\documentclass[12pt,preprint]{aastex}
\usepackage{graphicx}

\def\lta{{\>\rlap{\raise2pt\hbox{$<$}}\lower3pt\hbox{$\sim$}\>}}
\def\gta{{\>\rlap{\raise2pt\hbox{$>$}}\lower3pt\hbox{$\sim$}\>}}
\newcommand{\ea}{et al.~}

\newcommand{\ser}{$r^{\frac{1}{n}}$}
\newcommand{\hi}{H$\,${\scriptsize I}}

\shorttitle{Old and young late-type bulges}
\shortauthors{Carollo et al.}

\begin{document}

\title{Old and young bulges in late-type disk galaxies\footnote{Based on observations with the NASA/ESA {\em Hubble Space Telescope}, obtained at the Space Telescope Science
Institute, which is operated by AURA Inc, under NASA contract NAS
5-26555.}}

\author{
C. M. Carollo\altaffilmark{1}, C. Scarlata\altaffilmark{1},
  M. Stiavelli\altaffilmark{2},   R.F.G.  Wyse\altaffilmark{3}, 
  L. Mayer\altaffilmark{1}
  }

\altaffiltext{1}{ETH Zurich, Physics Department, CH-8033 Switzerland}
\altaffiltext{2}{Space Telescope Science Institute, Baltimore, MD 21218, USA}
\altaffiltext{3}{Department of Physics and Astronomy,  Johns Hopkins University, 3400 North Charles Street, Baltimore, MD 21218, USA}

\begin{abstract}
  We use HST ACS and NICMOS multi-band imaging to study the structure and the inner optical and
  near-infrared colors of a sample of nine late-type spirals. We use  a range of  population-synthesis-based star formation histories to interpret the observed $B-I$ and $I-H$ colors.
   We find: \\
$(1.)$ A correlation between  bulge and disks scale-lengths, and a correlation between the colors of the late-type bulges and those of the inner disks. Our data show a  trend for the bulges to be more metal-enriched than their surrounding disks, but otherwise no simple age-metallicity connection between these systems. This suggests a complex relationship between the star formation and metal production histories of late-type bulges and their surrounding disks; \\
 $(2.)$ A large range in  bulge color properties, indicating a large range in stellar population properties, and, in particular, in stellar ages. Specifically, in about a half of the late-type bulges in our sample the bulk of the stellar mass was produced  rather recently. This indicates that, in a substantial fraction of the  $z=0$ disk-dominated bulged galaxies, bulge formation occurs after the formation/accretion of the disk;\\
 $(3)$ In about a half of the late-type bulges in our sample, however, the bulk of the stellar mass was produced at early epochs, as     is found  for the early-type, massive spheroids;  \\
 $(4.)$ Even  the  ``old"  late-type bulges  host a   significant    fraction of stellar mass in a young(er) component, indicating that, in contrast with the very massive bulges,  possibly all late-type bulges are rejuvenated at later times. This is similar to what is found for the Milky Way Bulge;\\
 $(5.)$ A correlation   for bulges   between stellar age and stellar mass, in the   sense that     more massive late-type bulges   are   older than less massive late-type bulges.  Since the   overall   galaxy luminosity (mass) also correlates  with the bulge luminosity (mass), it appears that the galaxy mass  regulates not only what fraction of itself ends up in the bulge component, but also  ``when"     bulge formation takes place.\\

Our result extends to the smallest bulge mass scales --bordering with the typical masses of  nuclear star cluster-- previous  findings on the ages of stellar spheroids. The variety,  and the likely multi-burst nature,  of the star formation histories of    late-type bulges could be naturally explained by several processes   contributing,  at different epochs and on different time-scales, to the formation of  low-mass spheroids. On the other hand, the scalings between bulge stellar age and bulge/galaxy mass, and between bulge and disk scale lengths,  hint at similar processes   for all,   and suggest that   late-type   bulges of all (masses and) stellar ages result from the internal evolution of the parent disks. We show that  dynamical friction of massive clumps in gas-rich disks is  also, together with well-known bar-driven mechanisms, a  plausible disk-driven mode for the formation of late-type bulges, especially for those that are older than their surrounding disks. If disk evolutionary processes are   indeed   responsible for the formation of the entire family of late-type bulges, CDM simulations need     to produce a similar number of initially bulgeless disks in addition to the disk galaxies that are observed to be bulgeless at $z=0$.
 \end{abstract}


\section{Introduction} 

The late-type bulges of disk-dominated galaxies differ
significantly from merger remnants: They show disk-like cold kinematics (Kormendy 1993),
and Sersic profiles with $n-$values substantially smaller than the $n=4$ deVaucouleur 
value typical of an elliptical galaxy, and very
close to the $n=1$ that describes a disk-like, exponential profile
(see e.g., Kormendy \& Kennicutt 2004,  Wyse, Gilmore \& Franx 1997
and Carollo 2004 for reviews and references therein).

Whether externally- or internally-triggered, disk evolution is thought
to be likely responsible for the formation of these disk-like
bulges. Interactions with companion galaxies are for example
observed , statistically,  to induce a redistribution of gas within
a disk, which could lead to the formation of bulges (e.g. Kannappan et
al.\ 2004). Although quantitative predictions are scarce, the
accretion of small satellite galaxies onto a pre-existing galaxy disk
could also   drive gas into the central regions (e.g.~Hernquist \& Mihos 1995) and contribute   to the formation of a central bulge.  Within
a disk, a fire-hose (or buckling) instability can scatter the stars
originally in a stellar bar above the plane of the disk, into what
resembles a bulge-like structure (Raha et al.\ 1991; Debattista et
al.\ 2004; 2006).  A stellar bar can also be transformed into a
three-dimensional bulge-like system by the accumulation of   a large enough mass concentrated   in the
center, which can disrupt the regular orbits supporting the bar
(Norman et al.\ 1996; the most recent simulations however indicate
that a rather large central mass is required for the process to be
effective; see Shen \& Sellwood 2004).  Bulge formation via bar
instability is furthermore indirectly supported by arguments
  relating a disk surface density threshold to the onset of   AGN activity (Wyse
2004). Since several secular process could   induce   gas flows towards
the galaxy center, the resulting bulges could be even younger than the
surrounding disks.  Phase space density arguments favor dissipative
mechanisms to form bulges at late times (Wyse 1998), although
collisionless disk evolution does not violate phase-space density
constraints if it occurs at early times (Aliva-Reese et al. 2005).
Numerical simulations of bar-driven disk evolution with and without
gas show that the resulting bulges are not only structurally, but also
dynamically ,  similar to the bulges of disk-dominated galaxies (e.g.,
Debattista et al.\ 2004; 2006).  If     late-type bulges were the
outcome of such     processes acting {\it after\/} the formation of the disk,    about 30-40\%
of today's galaxies  -- that percentage being late-type disk galaxies in the local Universe -- should have been  born as pure (bulgeless) disks. 
This would be   an important constraint on    galaxy
formation models.

  The breaking of   degeneracies amongst possible formation processes requires
both quantitative predictions on the statistical properties of bulges
that form through different mechanisms, and observational diagnostics
capable of breaking such degeneracies. Qualitatively, one might expect
that e.g., thick disks with properties related   to those of    the bulge
(unfortunately measurable only in edge-on systems) might be
a signature of bulge formation by satellite accretion   (e.g., for the Milky Way, see Wyse 2001), while e.g.,
internal secular evolution of the parent disks might be a more natural
explanation for correlations between the stellar populations and the
scale lengths of bulges and disks.  Studying the properties of
late-type bulges is however rather difficult on account of their small
sizes, which are typically barely resolved in seeing-blurred,
ground-based images. The spatial resolution of ground-based images is
  non-optimal  to mask out dust features and star forming regions, which
affect the measurements of the properties of the underlying stellar
populations, and to disentangle   cleanly   from each other the light
contributions arising from different galactic subcomponents.
In contrast, the high spatial resolution achievable with the Hubble Space Telescope (HST) allows the exploration of     nearby disk galaxies on     $\sim$10pc scales.  
Our HST WFPC2 and NICMOS survey of $\sim100$ disk galaxies (Carollo et al.\ 1997, 1998, 2001, 2002; Carollo 1999; Seigar et al.\ 2002) has for example unveiled that about $70\%$ of these systems host  massive star clusters in their isophotal centers, whose light contribution needs to be masked out   prior to the analysis of   the bulge and disk components (see Appendix A).

This and a companion  papers (Paper 1 and Paper 2, respectively) continue our previous observational efforts to investigate the properties of the late-type galaxy population with the HST, and specifically present new data which set additional  constraints on the structural and stellar population properties of  these systems. In this Paper 1 we focus  on the late-type bulges, and on  their relation with the properties of their host galaxies, with emphasis on their stellar population properties. While generally   less sensitive to   the  mass assembly process,   photometric   stellar population diagnostics are key to test the epoch and timescales for the formation of the stars that are locked in bound structures. In Carollo et al.\ (2001) we showed that the $V-H$ colors
of late-type bulges are on average bluer by $\sim 0.5$ mag than the
colors typical of the massive spheroids.  We have interpreted these
blue colors as indicative that many late-type bulges are made of
stellar populations which are, on average, significantly younger than
those of the massive early-type bulges and of the elliptical
galaxies.   A single color is however notoriously degenerate towards stellar age,
metallicity, dust and   contamination by emission lines.  

To characterize more accurately the properties, and in particular the 
metallicities and stellar ages of the central sub-galactic components of late-type disk galaxies,  we have followed up our previous HST survey by acquiring new ACS data in $U$, $B$ and $I$ for  eleven galaxies in our original WFPC2+NICMOS sample,   nine of which have   bulge surface brightness distributions  well fitted by an almost exponential profile; the latter nine galaxies are thus those that we discuss in this first paper. Only the $B$ and $I$ images have enough signal-to-noise ratio in the continuum to allow the study of the bulges and inner disks. In this Paper 1 we therefore focus on combinations of  data in   $B$, $I$ and $H$ filters  (the latter available from our previous survey); these  have different sensitivities to effects of dust, metallicity and recent star formation,  and thus help to break the degeneracy in these parameters (e.g., Knapen et al.\ 1995).   

This paper is structured as follow. Section 2 and Appendix 1
provide information on the sample and the data reduction and analysis. 
In the Appendix we also present a discussion   of   the possible effects of
dust reddening on our analysis, and   tabulate   the structural parameters for the
galaxies (as obtained through ellipse fitting). The surface brightness radial profiles are made available 
in electronic tabular form. In Section 3 we present and discuss the (structural and) stellar population properties of the bulges and inner disks, their  relationships,  and their dependence on the properties of the host galaxies. In Section 4 we discuss possible implications of our results for the formation of the local disk galaxy population.  We summarize our main conclusions in Section 5.  
Throughout the paper we adopt   { $H_0=65$~km~s$^{-1}$~Mpc$^{-1}$}.  

We postpone the investigation --which we started in Carollo 1999 and Carollo et al.\  2001,2002-- of  the properties and scaling laws of the nuclear star clusters,  and their relation with the host bulges and disks, to Paper 2 (Carollo et al.\ 2006, in preparation). 

\section{The sample and  data analysis}
\label{sec:basicdata}

The nine galaxies of the present sample were selected from our  WFPC2 and NICMOS survey.  All
galaxies have $v<2500$kms$^{-1}$ and inclination $i<75^o$, to avoid strong obscuration of the
nucleus by the disk.  Table~\ref{tbl-1} lists their
coordinates and some of their global properties.
The nine targets are rather typical late-type disk galaxies in terms of bulge Sersic index (i.e., their bulges have almost-exponential light profiles with $n\le$2), disk scalelengths, rotational velocities, densities, and total magnitudes; on the other hand, they
probe the small-size end of the bulge population, which is not easy to investigate at ground-based resolution (see Appendix A.1 for further details). 

The galaxies were observed with the ACS-WFC in   each of   the F435W ($B$) and F814W ($I$) filters (and with the F330W filter in the ACS-HR channel - but the S/N of these  UV images is too low for studying the diffuse stellar components; the UV data are thus not considered in the present discussion). 
The standard ACS pipeline was used to perform the basic data
reduction, followed by   cosmic-ray rejection, correction for hot pixels   and sky
subtraction. Image alignment was checked before combining the
individual images in each filter; we also checked the alignment between  images in the two
different filters.  A photometric calibration was applied to convert
the instrumental magnitudes into VEGA magnitudes. These were
corrected for Galactic extinction following Schlegel, Finkbeiner \&
Davis (1988).

For each galaxy, we computed radial surface brightness profiles in $B$
and $I$ by fitting ellipses to the final images with the IRAF\footnote{IRAF is distributed by the National Optical Astronomy Observatories, which is operated by the Association of Universities for Research in Astronomy, Inc. (AURA) under cooperative agreement with the National Science Foundation.} program
ELLIPSE.   Star-forming   regions/knots and dust lanes and patches were
masked out before performing the fits. The profiles are plotted and
reported in electronic tabular form in the Appendix. For easy
reference, the $H$ band NICMOS profiles of Seigar et al.\ (2002), which are used
in the current analysis, are also given in the Appendix (augmented in
radial extent, when possible, using the 2 Micron All Sky Survey 
data (2MASS; Jarrett 2004).

The   $I-$~band   surface brightness profiles were  derived  with
a PSF--convolved, Sersic bulge plus exponential disk model.  The
innermost radial points in the profiles were excluded when performing
the analytical fits for eight of the nine galaxies, as the light
emission is there dominated by the nuclear star clusters which have
been previously identified in the galaxy centers. In ESO~499G37, the closest  galaxy in our sample
(at a nominal distance of only 14Mpc),  the size/mass of the well-resolved "bulge" reaches the typical scales that are observed for  the nuclear star clusters  in the other more distant galaxies in the sample. In the latter galaxies, the nuclear clusters are found in addition to a bulge component; in ESO~499G37, the cluster-sized bulge is the only central structure detected in addition to the disk.  It is an interesting question whether ESO~499G37 is a  bulge-less disk with a nuclear star cluster (for which we  resolve its spatial structure), or  a disk galaxy with a small bulge and no central star cluster. In Paper 2 we show that, in terms of surface mass density versus stellar mass, the central structure in this galaxy can be considered as a "transition object" between bulges and nuclear star clusters. We have opted for including the cluster-sized "bulge" of ESO~499G37 in the current analysis, and treating this structure similarly to the other bulges. With the "bulge" of ESO~499G37 we extend the study of bulges of late-type galaxies down  to the $\sim 10^6$M$_\odot$ scales; we will discuss further in Paper 2 the relationship between nuclear clusters and host bulges, and what can be learned about this relationship from the special case of ESO~499G37. 

Tables~\ref{tbl-2a} and ~\ref{tbl-2b} list the main model parameters
derived for each of the galaxies from the $I$ images. Since the
$B$-band surface brightness profiles   show more features   than the
$I$-band profiles, and the $H$-band profiles are typically noisier and
less extended than the $I$-band profiles, we used the structural
parameters of the $I$-band   best-fit   analytical models to derive the
$B$ and $H$ bulge total/central magnitudes, which are also listed in
the Tables (see also Appendix A.2).  We stress that this   is equivalent    to
setting to zero,  by construction,  any possible color gradient   in the bulges
(which are thus not addressed in this paper); these are expected to be 
smaller than the color gradients between bulges and disks (Kormendy \& Kennicutt 2004), 
which we measured after accounting for the differences in the PSFs of the
various passbands. In Table~\ref{tbl-col} we list the     
differences between the bulge colors, derived from the photometric
decompositions as discussed above, and the colors of the disks, as
estimated at a galactocentric distance of 5 bulge half-light
radii. Since the bulges are best fit by close-to-exponential profiles,
and the half-light radius for an exponential is equal to 1.67
scale-lengths, this thus corresponds to measuring the disk colors at
more than 8 bulge scale-lengths, i.e., at a distance equal to a large
fraction of the disk scale length, and thus well into the
disk-dominated regions.

\section{Stellar populations in the centers of late-type disk galaxies}

\subsection{Two caveats}

\begin{figure}
\begin{center}
\includegraphics[scale=0.6]{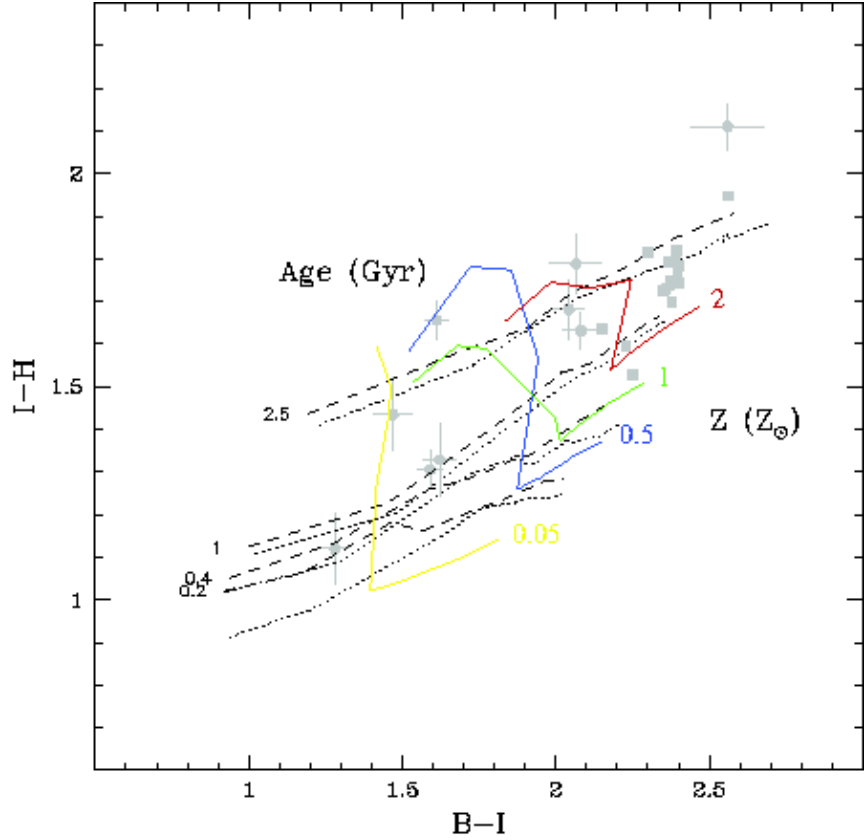}
\caption{ The $I-H$ versus $B-I$ color-color plane, with   grids of
SSP models from Bruzual \& Charlot (2001, BC01; dashed black lines),
Bruzual \& Charlot (2003, BC03; dotted black lines), and Maraston
(1998; color lines). Each line is an isometallicity track   with ages
  ranging from 0.5 Gyr (left) to 12 Gyr (right).  In the Maraston
models,   different   colors represent different metallicities, indicated in solar
units on the right of each track. The metallicities of the BC01 models
are indicated in solar units on the left of each track; these are
identical for the BC03 tracks, which appear slightly   ``  down-shifted"
with respect to the BC01 tracks. For reference, the grey symbols
indicate the colors of the late-type bulges of our study (circles) and
the massive bulges of Peletier et al.\ (1999; squares).
\label{fig:testmods}
}
\end{center}
\end{figure}

We stress two issues that in principle might have some impact on our conclusions, before presenting our results on the colors and thus stellar populations in the centers of the late-type galaxies of our   sample:

\begin{itemize}

\item Even with the availability of several passbands, disentangling the effects of stellar ages, metallicities, dust and emission lines remains a non-trivial task.  
The high spatial resolution that is achieved with the ACS allows us to easily mask out
sharp dust features or compact sources such as regions of very recent
star formation.  This allows us to measure reliable colors of the
diffuse, underlying stellar populations.  A smoothly distributed dust
distribution ,  however,    would remain undetected in our analysis (as in any other
study based on broad-band colors).  We discuss in the Appendix what
the impact of any such a diffuse dust component would be on our main
results.   This   impact is mainly to render the absolute stellar
population ages and metallicities rather uncertain.  Furthermore, if galaxies with larger stellar mass and thus higher metallicity had more dust (as suggested by, e.g., Tully et al.\  1998), the ages of their stellar populations could be systematically overestimated, a bias that could affect the comparison between the stellar population properties of galaxies with very different metal content and stellar masses. 

\item There are substantial differences between different population
synthesis models, which hamper the determination of absolute ages and
metallicities of integrated stellar populations. As an example, in
Figure~\ref{fig:testmods} we show, on the $I-H$ versus $B-I$
color-color plane that we use below as our main disgnostic, a grid of
{\it simple stellar population} (SSP) models extracted from the
Bruzual \& Charlot compilation (2001, hereafter BC01; web publication;
dashed black lines). The almost-horizontal, dashed lines are
isometallicity tracks with metallicity ranging from Z=2.5$Z_{\odot}$
to $Z=0.2Z_{\odot}$ from top to bottom; for each metallicity, ages
range from 0.5 to 12 Gyrs from left to right. The grid is obtained
using a Salpeter (1955) stellar initial mass function (IMF) with a
lower and upper mass cut-offs of 0.1$M_{\odot}$ and 100$M_{\odot}$,
respectively.  The use of a Salpeter IMF all the way to the brown dwarf limit 
is a  choice of  convenience that does not significantly affect our "comparative" results.
A similar grid is reproduced for the models of Bruzual
\& Charlot (2003, hereafter BC03; black dotted lines). Finally, an
equivalent grid is shown using the Maraston (1998) models (hereafter
M98; color solid lines). Their metallicities are 2Z$_\odot$,
$Z_\odot$, 0.5$Z_\odot$ and 0.05$Z_\odot$ from top to bottom, and the
stellar ages range, as before, between 0.5 and 12 Gyr from left to
right.  The reason why we choose to compare these specific models is
that there are substantial technical differences between them, e.g.,
the BC01 and BC03 models are based on an   ``isochrone  synthesis" approach
and the M98 models are based on the   ``fuel  consumption" approach
introduced by Renzini (1981; see Maraston 1998 for an extensive
discussion). For our discussion, however, the most significant
difference between the Bruzual \& Charlot models and the M98 models is
a different recipe for the inclusion of a thermally-pulsing
Asymptotic Giant Branch, which substantially changes the
SED properties of intermediate-age stellar populations. 

 This is evident from the Figure, which shows substantially
different predictions for the colors of stellar populations of about 2
Gyr or younger.  For reference, we also plot on the diagram the
measurements for our late-type bulges (grey circles) and for the
massive early-type bulges of Peletier et al.\ 1999 (grey squares; see
below for details on the measurements).  From the Figure, it is clear
that there is a large uncertainty, associated with the models, in
deriving absolute ages and, in particular, absolute
metallicities. However, the relative ranking remains robust,
especially for the stellar ages.

In the analysis that we present below, we adopt the BC01 models to set quantitative constraints on the stellar population properties of the late-type bulges and their surrounding disks in comparison with each other and with the properties of the more massive early-type bulges.

\end{itemize}

While being aware of the caveats discussed above,  we trust that the main results of our work, due to the size of the effects that we discuss and to the comparative nature of our statements,  are not strongly affected by either an invisible diffuse dust component or the particular choice of population synthesis models.

\subsection{The bulges}

\subsubsection{The simplest analysis: Comparison with SSP models}

The simplest assumption concerning the stellar population properties
of a stellar structure is that this is made of stars with identical
ages and metallicities. While this is unlikely a good
approximation of reality, most studies perform such a comparison
between the observed colors of integrated stellar population and the
SPP models describing such a scenario.
The $I-H$-$B-I$ grid of  BC01 SSP models of Figure~\ref{fig:testmods}  is again reproduced in Figure~\ref{fig:MIvsIH}, with overplotted  the measured central colors of the low-Sersic-index, late-type bulges of our study (blue solid circles), and isochrones of ages ranging from 0.5 to
12 Gyr (almost-vertical dashed lines).
The bulge colors are computed
using the Sersic fits to the observed bulge brightness radial profiles (adopting the $I$-band analytical best fit structural parameters for all passbands).
Assuming that the real disks are exponential, as used in our
analytical descriptions, this should eliminate any contribution to the
color estimates arising from the disk component; if the 
stellar disks were more concentrated than an exponential profile,
subtracting the latter would leave a residual contamination
from the disk component to the light attributed to the bulge, but would nonetheless
minimize such contamination. 
The fact that a significant fraction of nearby   stellar   disks has
a perfectly-exponential profile from the many kpc scales down to the
innermost point sampled by the HST  (B\"oker et al 2003) suggests
that  an exponential profile for the underlying disk components
is most likely an accurate description for real disks. 

In Figure~\ref{fig:MIvsIH} the early-type, massive bulges studied by
Peletier et al.\ (1999) are also plotted for
comparison (black squares). Their bulge colors were derived on
the basis of NICMOS-F160W, WFPC2-F450W, and WFPC2-F814W images; the
conversion to our filter set was derived by using a   10~Gyr  SSP of
solar metallicity, and   is:  $(B-I)=(B-I)_{\rm Pel} + 0.3$;
$(I-H)=(I-H)_{\rm Pel} -0.02$.  No disk subtraction was attempted in
these measurements. However, in early-type disk galaxies, the bulge strongly
dominates the light budget in the galaxy centers; therefore, the disk
contribution in the Peletier et al.\  measurements is almost negligible.
The colors that we adopt  for these massive bulges are those measured at one half-light radii from the galaxy centers, as the nuclear colors are affected by the presence of dust, nuclear point sources and/or stellar clusters.  As visible in Figure 2 of Peletier et al.\, the color gradients of these systems outside of these affected nuclear regions are much smaller than the effects that we discuss in this paper; therefore, we trust the color comparisons between the late-type bulges and  the massive bulges to be robust.

 The colors of the massive, early-type bulges  are well
reproduced by old ($\sim 9 \pm 2$ Gyrs) and metal-rich stellar
populations, in agreement with the original analysis of Peletier and collaborators.
The two samples that we are comparing clearly cover, on average,
different regions of the $I-H$-$B-I$ color-color diagram, suggesting different 
stellar ages and metallicities. While the
early-type, massive bulges are concentrated in a region of parameter
space corresponding to old and metal-rich stellar populations, the
late-type bulges cover a wide range of observed $B-I$ and $I-H$
colors. Using the SSP models as a benchmark to infer from the observed colors
the stellar population properties of the late-type bulges,  stellar
population ages ranging from less than 1 Gyr up to
$\sim 3$ Gyrs and a large spread in metallicity   are   indicated
for these systems. This comparison, at face value, would support
previous work   concluding    that     late-type bulges are, on average, 
younger and more metal poor than   massive, early-type bulges.

A few late-type bulges lie, on the color-color diagram, well outside
the region covered by the grid of SSP models, requiring
  dust-reddening   effects in order to reconciliate the observed colors
with SSP models.  The reddening vectors for an absorption of $A_V=0.5$
mag are shown for the triplex model of Disney et al. (1989, {\it red})
and for an idealized uniform foreground screen model ({\it black}) in
the upper--left quadrant of the Figure. The two reddening vectors run
almost parallel to each other, and indicate that reddening effects
could likely produce the observed discrepancy between observations and the SSP
models. If dust affects the observed colors of the late-type bulges,  
it would most likely work in the direction of reddening their intrinsically bluer colors, and it would thus
 imply even younger and less metal-enriched stellar populations for these systems.  
 In Table~\ref{tbl-4} we summarize the stellar population properties of the bulges in our sample
that are derived from  the comparison with the SSP models.

The SSP models are however most likely inadequate to describe the star
formation histories (SFHs) of the late-type bulges. In particular,
they do not provide information as to whether the colors of the
late-type bulges arise from mostly old stellar structures that are
slightly rejuvenated by minor (in terms of stellar mass) recent star
formation events, or whether it is the bulk of the stellar mass in the
late-type bulges that is relatively young - and younger than the
stellar populations of the massive, early-type spheroids.  We
therefore adopt below two different SFHs for the late-type bulges, and
use these more realistic models to set constraints   on  the age of
the bulk of the stars in these systems.    For only one galaxy,
NGC~3259, do the arguments favoring the presence of dust remain
supported even when using more realistic star formation
histories; for all other galaxies in our sample, the composite stellar population 
model grids can explain the observed colors of
their bulges.  It might be worth noting that NGC~3259 is the only galaxy in our sample with a {\it  spatially unresolved} central compact "object" (at HST resolution; see Carollo et al.\ 2002), in contrast with all other galaxies in which, at similar distance, the distinct photometric nucleus is resolved into a nuclear star cluster.

\subsubsection{Comparison with   Composite Stellar Population    models}

The simplest approach to   quantify the presence of   an old stellar population
component in relatively blue stellar systems ,    such as the late-type bulges of our study ,   
is to   adopt a model in which    their stars were born in two distinct starbursts events,
the first taking place early-on in the life of the universe, and the second one
occurring at a later time.  

We created such  {\it   composite    stellar population} (CSP) models by adding, to 12 Gyr old SSP models, a  fraction $f$  of the total stellar mass in young stars. We considered young components with
ages of $0.1, 1, 3, 5$, and 8 Gyrs, with metallicities ranging from
0.2$Z_{\odot}$ to 2.5$Z_{\odot}$. The fraction $f$ of young stars
ranged from $0\%$, reproducing a   12Gyr-old   SSP, up to $100\%$,
reproducing a SSP model of age equal to that of the young component.

We show in Figure~\ref{fig:COMPOSIT} the $I-H$ vs $B-I$ color-color
diagram   of our sample bulges,   with overplotted the grid of CSP
models for three different ages of the young stellar component, namely
  100 Myr (top), 1 Gyr (middle) and 3 Gyr   (bottom).  In each panel
of Figure~\ref{fig:COMPOSIT} only the mass fraction and metallicity of
the young stellar component are varied.  For clarity, only tracks
corresponding to $0.2, 1.0,$ and 2.5 $Z_{\odot}$ are plotted for the
young stellar component (cyan, black and magenta tracks,
respectively); the same metallicity range is more densely sampled for
the old stellar population (black labels to the right of the
almost-horizontal isometallicity tracks).

As evident in Figure~\ref{fig:COMPOSIT},  in principle
 a small ($<<10\%$) fraction of the stellar mass with a stellar
age of order 100Myr  could reproduce the observed blue
colors of the late-type bulges.  However, the relaxation time for these systems is typically of order several hundred million years; therefore,   stellar populations that are significantly younger than such timescale would possibly be not dynamically-relaxed, and would likely appear as clumped star forming regions. Thanks to the resolution achievable 
with the HST, we were able in our analysis
to mask out bright knots of recent star formation. We
therefore consider this interpretation for the blue
colors of the underlying diffuse stellar populations of the
late-type bulges as unlikely. 

At the other extreme,
 should the late-type bulges host a   12~Gyr-old   
underlying stellar population, Figure~\ref{fig:COMPOSIT} shows that,
consistently with the  stellar ages inferred from the SSP analysis of
Figure~\ref{fig:MIvsIH}, a stellar component substantially
younger than   3~Gyr  would be required to explain their blue colors, as none
of the late-type bulges overlaps with the theoretical models in the
case of a   3~Gyr-old   ``young'' stellar component.

We therefore adopt the 1 Gyr-old  stellar population 
as a benchmark to discuss the properties of the late-type bulges based 
on the CSP models. If the most recent burst of star formation in the late-type bulges
occurs about 1 Gyr ago,  a substantial  ($> 20$\%) fraction of the stellar mass in such young
stars is required   to explain the blue colors of about a half of the  late-type bulges in our sample. 
In contrast, however,  the other half   of our sample of   late-type bulges are  well
described by CSP models   with    less than about 10\% of   the   mass in 1Gyr-old
stars (where 10\% is the upper limit derived by considering the most
metal-rich tracks for the 1Gyr-old stellar component).  
In fact, in about a half of the late-type bulges in our sample
 the colors are consistent with their stellar mass budget being dominated 
 by a very old stellar population. 
 
 This  variety  of stellar population properties unveiled 
 in the late-type bulges highlights   the    complexity in the 
star formation histories of these stellar systems.

\subsubsection{Comparison with exponential star formation
history  models}

The second approach that we use to investigate the presence of an old
stellar component in the late-type bulges allows us to explore the
effects on the colors of an extended period of continuous star
formation (as opposed to the   ``two bursts" model described above).
Here we assume that the late-type bulges have SFHs that are well
described by an exponentially decreasing star formation rate (SFR).
The bulges SFRs are parametrized by the $e-$folding time of star
formation, $\tau$.  We label these models {\it exponential star formation
history} (ESFH) models.  Under the assumption that the IMF does not
depend on time, and that the stellar metallicity remains constant with
time, the integrated spectrum $F_{\lambda}(t)$ at time $t$ of a
stellar population characterized by a star formation rate $\psi(t)$
can be written as:

\begin{equation}
F_{\lambda}(t)=\int_{0}^t \psi(t-t')\,f_{\lambda}(t'){\rm d}t',
\end{equation}

\noindent where $f_{\lambda}(t')$ is the time-evolving spectrum of a
single stellar population of age $t'$, and $\psi(t)\propto
e^{-t/\tau}$ is the SFR of the galaxy as a function of time.  Assuming
that all bulges started to form stars   at a fixed time,    variations in their
observed colors depend on their e-folding time $\tau$ and on the
metallicity of their stellar population.  Their SFHs are thus
described by these two parameters. 
A SFH--weighted average age
($<A>$) of the   composite   stellar population can be  defined as (see, e.g., MacArthur et al. 2004):
\begin{equation}
<A>= T- \frac{\int_0^Tt\psi(t){\rm d}t}{\int_0^T\psi(t){\rm d}t},
\end{equation}
where $T$ is the age of the oldest stars, namely when the star
formation started, which we assumed equal to   12~Gyr ago   for all models.
In this formulation, the metallicity is kept constant during   the entire    star formation history, which 
 is of course a rather coarse approximation. This has the advantage however of avoiding 
 introducing further assumptions on specific chemical evolution models.  

In Figure~\ref{fig:ESFH} we show the $I-H$ vs $B-I$ color-color
diagram of Figure~\ref{fig:MIvsIH} ,   with overplotted the theoretical
grid computed assuming these ESFH models.  The e-folding time of star
formation is used to label the almost-vertical isochrone curves.  The
metallicity of the different models varies from 0.2 to 2.5
$Z_{\odot}$, and it is indicated on the right side of the grid to
label the almost-horizontal isometallicity curves.
Table~\ref{tbl-5} summarizes the stellar population properties of the late-type
bulges that are derived from  the comparison with the ESFH models.

Figure~\ref{fig:comppel}  visualizes the main difference
in average stellar population properties between the early-type
massive bulges of Peletier et al.\ (1999; grey shaded area), and the
late-type bulges in our sample (histograms).  In contrast with the old
and broadly-speaking coeval, highly-enriched massive bulges, the
bulges of disk-dominated galaxies are indeed a much less coherent
population, covering a wide range in stellar metallicities and
ages.

In Figure~\ref{fig:sfhhistmass} we show the SFH (i.e., SFR as a function of time)
that better reproduces the observed colors for each late-type bulge in
our sample.  In this figure, zero represent the redshift z=0 point.  
Some of the late-type bulges  , such as 
NGC~2758,   are well reproduced by an almost-constant SFH. However,
other late-type bulges ,  such as  ESO~498G5,   are better described
by a SFH which peaks several   Gyr   ago, suggesting that a   substantial   
fraction of stellar mass in these systems is in old stars.

Using the derived SFHs, we  quantify the amount of stellar mass in 
the late-type bulges of our sample  that formed  in the first and in the last three billion years, respectively. In particular, Figure~\ref{fig:sfhhistmass} shows, for the nine late-type
bulges of our sample, the distribution of mass fraction which formed
between 12 and   9~Gyr   ago (red histogram), and the stellar mass which
formed  from   3~Gyr   ago to today (blue histogram; see also Table~\ref{tbl-5}).

About a half of the  late-type bulges in our sample  formed a non-negligible fraction of their
stellar mass  in the last   3~Gyr. Since the disks will have been in place by then, 
this supports that
secular evolution processes play an important role in building up
additional stellar mass in the centers of the   galaxy    after the assembly
of the disk itself.  However, in the remaining  late-type bulges,
the colors are consistent with these bulges  having formed more than
$50\%$ of their stellar mass   more   than   9~Gyr   ago, similar to   the conclusions concerning 
massive, early-type bulges and  elliptical galaxies   
(e.g., Thomas et al.\ 2005; Thomas \& Davies 2006).

Figure~\ref{fig:spmass} shows the   dependences  of the   derived   stellar ages and
metallicities of our late-type bulges,   obtained   from their best-fit
ESFH models, on the stellar mass of the bulges, obtained from their
$H$ absolute magnitude. We report no significant dependence
between the metallicity of the   model  that provide the best   fit   to
the   color of a late-type bulge, and the bulge stellar mass (although, as
suggested by the referee, the assumption of
a constant metallicity in our models may contribute to dilute a correlation
between bulge stellar mass and metallicity, if this were present).
Despite a large scatter in stellar age, we find instead a clear
  correlation   between bulge mass and age, in the direction that more massive
late-type bulges are on average older stellar systems than the less
massive ones. This result has previously been presented concerning the
comparison between massive early-type bulges and less massive
late-type bulges (e.g., Carollo et al.\ 2001; Thomas \& Davies
2006). However, we show here that such a trend remains valid {\it
within} the family of late-type disk galaxies, i.e., the relatively
more massive late-type bulges are older than the smallest members of
their own family. Our statistics is inadequate to explore whether
important factors such as environment induce or contribute to this
trend.  The $H$ luminosities -good proxy for (stellar) masses- of
bulges and host galaxies are found to correlate however quite well,
down to the faintest luminosities probed by our late-type bulges (see
  left    panel of Figure~\ref{fig:bulgetot}). It would thus appear that
the galaxy mass is influential in establishing not only what fraction
of itself ends up in a central bulge component, but also at what epoch
this happens during the lifetime of the galaxy.

\subsection{The inner disks}

The analysis of ground-based imaging data has shown that the massive
bulges and their inner disks have
similar stellar populations  (Peletier \& Balcells 1996).
The photometric decompositions that we have performed on the ACS surface brightness
profiles allow us to compare the colors of the (smaller) late-type bulges with
those of their surrounding disks.

In the left panel of Figure~\ref{fig:bulgeinnerdisk} we plot the $B-I$
color of the late-type bulges (small symbols) and
of their surrounding disks (large symbols). The $B-I$ color of the
disks are those of the analytic disk components, as measured at five
half-light bulge radii from the galaxy centers (see also Figure~\ref{fig:colorprofiles}, 
which shows the $B-I$ and $I-H$ color profiles). Different symbols
refer to different galaxies. The vertical axis is used to arbitrarily
shift the points relative to each galaxy, to avoid overlap.  Only as a
benchmark, we indicate on the top axis the stellar age that is
associated with the corresponding $B-I$ color that is reported on the bottom axis, as
derived from SSP models of solar metallicity.  On average, the
majority of the late-type bulges have redder $B-I$ colors than their
surrounding disk. The bulge-to-disk $B-I$ color difference varies
significantly from galaxy to galaxy, ranging from $\sim -0.2$ mag in
NGC~2082, the only galaxy in our sample with a bulge bluer in $B-I$ than the
disk, to $\sim 0.5$ magnitudes.  The average value of $B-I$ color
difference between the bulge and the disk is $\Delta (B-I)=0.27\pm
0.20$.

The right panel of Figure~\ref{fig:bulgeinnerdisk} shows the $B-I$
color of the bulge plotted versus the $B-I$ of the disk. The red solid
line is the best fit  to the observed values [$(B-I)_{\rm
  bulge}=(0.9 \pm 0.2)(B-I)_{\rm disk}+0.5\pm 0.3$].  As a benchmark to illustrate 
  the effects of reddening on this relationship, the
arrows indicate the change in the bulges' $B-I$ colors that is required 
 in order to bring the observed points to the
margin of the theoretical SSP grid (see Figure~\ref{fig:MIvsIH} and Table~\ref{tbl-4}); the dashed red line represents the best
fit derived to the relation as derived by using   these estimate of   the de-reddened colors
[$(B-I)_{\rm bulge}=(0.5 \pm 0.2)(B-I)_{\rm disk}+0.8\pm 0.3$].  The use of the CSP or ESFH models would require reddening only for one galaxy, NGC 3259, in an amount similar to the one derived from the SSP models. Thus, the adoption of the reddening  values derived from the SSP grid  allows us  to show that the correlation between bulge and inner  disk $B-I$ colors is robust towards the application of  some reasonable reddening corrections.
  
The observed correlation between the $B-I$ colors of  bulges and  inner
disks in late-type galaxies supports previous suggestions, also based on a single broad-band color,
 of a connection between the SFHs of these systems.
 This connection, however, appears to  be more complex than previously suggested.  
 It is  in fact difficult to interpret a single (e.g., the $B-I$)  color in terms of stellar population properties: the point is made by e.g., NGC~2082, which, in contrast with what is observed in $B-I$, in $I-H$ has a bulge that is redder than the inner disk (Figure~\ref{fig:colorprofiles}).
When more than one color becomes available, it is clear that the link
between the stellar population properties of bulges and inner disks in
late-type spirals is not a trivial one. This is shown on the $I-H$-$B-I$ plane
of Figure~\ref{fig:bulgeinnerdisk2}, where the arrows describe  the inner disk (end of arrows) and bulge (start of
arrows) colors for the five galaxies for which we have extended enough measurements in all passbands.   The bulges appear to be generally more metal-enriched than their surrounding disks; however,
the age-metallicity relationship between bulges and their inner disks
is rather complex and varies from galaxy to galaxy, in no systematic way.  For example, in NGC~2082, the bulge is more enriched but younger than the surrounding disk, suggesting the occurrence of gas inflow and the generation of new stars in the center of this galaxy.

This complex situation, also found by Moorthy \& Holtzman (2005) on the basis of dust-independent line strengths analysis, is at variance with
what is found for the color differences between bulges and inner disks
in early-type spirals, which are known to be not only small but also
rather homogeneous (see e.g., Balcells \& Peletier 1996). This can be seen from the
gray shaded area in the Figure, which shows the scatter around the
color differences between early-type bulges and their surrounding
disks. The data are taken from Peletier \& Balcells (1996), who published the
average color difference (bulge-disk ) $\Delta(B-R)=0.078 \pm 0.165$
and $\Delta(R-K)=0.016\pm 0.087$ of a sample of early-type spirals; we
have converted these colors on our $B-I$-$I-H$ diagram by adopting
$B-I=B-R+0.8$; and $I-H=R-K-1.1$.  These relations were computed using
SYNPHOT in STSDAS,  for a 12 Gyr old stellar population with solar
metallicity and   exponential   SFH with $\tau=2$Gyr (parameters  
that are adequate to describe the stellar populations of     massive bulges).

\section{Discussion: Late-type bulge formation through cosmic history}
\label{sec:discussion}

While it is widely accepted that massive bulges are   metal-rich  
and old stellar structures, it is often forgotten that a similar
statement is true for their {\it inner} disks.  The tight relation
discussed above between ages and metallicities of bulges and inner
disks in the early-type massive spirals indicates that the formation
of these galactic subcomponents is highly correlated. There are
several plausible scenarios to explain this, at least on a qualitative
level, e.g., the inner disks in massive spirals might be either the
high angular momentum left-overs of the mergers that formed the
massive bulges, or the remnants of early dense disks within which
instabilities led to the massive bulges (e.g., Avila-Reese et al.\
2005).  The large spread in stellar population properties unveiled in
the central regions of the late-type spirals, instead, and, in
particular, the spread in bulge stellar ages and the complex
relationship between bulge and inner disk population properties,
highlight more complex star formation histories for the late-type disk
galaxies.  An important question is whether the diverse star formation
histories of the late-type bulges and inner disks are due to a variety
of mechanisms acting over cosmic history, or to a single mechanism
capable of acting at different epochs, and possibly over different
timescales and with different efficiencies.  The combination of
stellar population and structural diagnostics can set some
constraints.

\subsection{Young late-type bulges}

First, our results indicate  that, in several late-type bulges,  a substantial fraction of the stellar mass has a stellar age that is comparable with, or even younger than, the surrounding inner disks:  In the universe,  bulge formation  --or at least bulge growth-- can take place together with (or even after)  the formation/accretion of the disk.   

To be more specific, we compare the observed color distribution of our
late-type bulges with the theoretical predictions of Bouwens et al.\
(1999).  This comparison does not aim at a quantitative interpretation
of the observed colors (as any model makes of course specific
assumptions on several key factors), but rather to illustrate some
global trends which can be trusted to be robust against variations in
the model details. The Bouwens et al.\ models assume a fiducial disk
evolution model that is constructed assuming that: {\it (i)} The
formation times are identically distributed as in Lacey \& Cole
(1993), with the exception that the halo formation time is set equal
to the time over which 25\% (instead of 50\%) of the final halo mass
is assembled. The formation time of galaxies of a given luminosity is
derived assuming a constant M/L ratio and the CDM power spectrum of
White \& Frenk (1991). {\it (ii)} Star formation in the disk starts at
the halo formation time, and has an $e-$folding time $\tau=3$ Gyr. The
latter is constrained by fitting the redshift $z=0$ color-magnitude
relationship.  {\it (iii)} The disk evolution in metallicity is given
by the standard equations by Tinsley (1980), with the yields tuned,
for each luminosity separately, to reproduce the observed   $z=0$   disk
metallicities.  {\it (iv)} Dust is described by a screen model with an
optical depth consistent with the observational constraints. {\it (v)}
The   stellar   disk is well described by an exponential profile with a central
surface brightness consistent with the Freeman (1970) law.

The bulge
component is then added assuming three different prescriptions for
bulge formation, namely: {\it (1.)} A   {\it secular evolution\/}   scenario
(SE), in which bulges form dissipatively, after the disks, due to
central star bursts generated by bar-driven inflow of gas into the
galaxy centers; {\it (2.)}  A   {\it simultaneous formation\/}   scenario
(SF), in which   the high angular momentum gas collapses to form   the
disk, while the low angular momentum gas, which sits closer to the
center of the dark halo, forms the bulge; and {\it (3.)}  An   {\it
early formation\/}   scenario (EF), in which bulges form through mergers
of disk galaxies, and the disk is then re-accreted at later times. For
the three different families of bulge-formation models, spectra are
computed using the Bruzual \& Charlot instantaneous-burst
metallicity-dependent spectral synthesis tables in order to determine
the colors of bulges and disks.

In Figure~\ref{fig:bouwens}, the solid, dotted and dashed lines
represent, respectively, the predicted age-sensitive $B-I$ color
distribution at z=0 for the SE, the SF and the EF models. The
theoretical $B-I$ colors have been obtained by converting the $B-R$
colors originally published by Bouwens et al.\ (1999; the conversion takes into account
 a small contribution that is introduced by the adoption of the 2001 version of the B\&C models).  Since the color
transformation depends on the SED of the object, we considered two
bracketing cases, i.e., a  10~Gyr-old   SSP with $Z=Z_{\odot}$, for
which we obtain $B-I = B-R +0.75$, and a 1 Gyr-old SSP with $Z=0.2
Z_{\odot}$,  which gives $B-I = B-R+0.65$. We adopt the
relation $B-I = B-I+0.7$, and associate   with   it an uncertainty of 0.1
magnitudes.

 In the Figure, the models are normalized to
the total number of galaxies in our sample, whose color distribution
is represented by the solid histograms. Following
Figure~\ref{fig:sfhhistmass}, we identify in Figure~\ref{fig:bouwens}
the bulges that formed more than 50\% of their stellar mass in the
first 3 Gyrs of their SFH ({\it red histogram}), and those that formed
less than 50\% of their mass in stars in the same period ({\it blue
  histogram}).  The Figure shows that the SF and EF
models of Bouwens et al.\ cannot reproduce bulge colors bluer than $B-I = 1.9 \pm 0.1$.
We have shaded this color-threshold to help visualizing the transition
between red colors, that can be explained with any of the models, and
blue colors that, within this set of models,  are inconsistent with a bulge formation which is
simultaneous or preceding the formation of the disks.  

In principle, it could be argued that late halo formation times might remain an alternative explanation for the young ages of these blue systems; however,  such late halo --and thus thin disk-- formation times are not favored by realistic disk-heating arguments, and by observations of the number density evolution up to redshifts $z\sim1$ of disk galaxies of different sizes (Lilly et al.\ 1998; Sargent et al.\ 2006).
Furthermore, different choices for the model parameters, e.g., a longer $\tau$ for the star formation in the disk,  can of course alter the exact value of the color-threshold. At least qualitatively, however,  it is reasonable to conclude that  only bulge formation via secular evolution of the
disks can explain the colors of the   bluest late-type bulges. 

Further indirect evidence in support of an internal origin rather than an external  trigger for  late bulge formation process that builds up the (blue) late-type  bulges possibly  comes from the correlation between 
the bulge half-light radii  and the disk scalelengths  which we show in the right panel of Figure~\ref{fig:bulgetot} for the nine galaxies of our sample (see also, e.g., Graham 2001). It is not clear whether a satellite-accretion scenario could reproduce such a correlation; in contrast, numerical simulations of  disk evolution driven by internal dynamical instabilities can reproduce this scaling law between the inner and outer regions of disk galaxies (e.g., Debattista et al.\ 2004).

\subsection{Old late-type bulges}

Not all late-type bulges in our sample argue however for a disk-first-bulge-after formation scenario.
In fact, several of them have colors and thus stellar populations which are consistent with the  more   ``classical"   bulge-first-disk-after picture for the formation of  bulges (e.g., Renzini 1999a,b; Renzini 2002). 
 The mass assembly history of these old bulges is  not strongly constrained by the observations. The old late-type bulges could well be the low-mass end of the merger remnants sequence.  Given however that old and young late-type bulges appear to be one family, at least in terms of the bulge-disk size scaling relation, it is tempting to consider the possibility that even the old late-type bulges are, as it is plausible for their younger counterparts, the outcome of (likely internal) disk evolution processes. 
 
Dissipationless evolution of early, dense stellar disks could be a
viable process for the formation of old late-type bulges: the expected
higher density of seed stellar   disks    in the early Universe provides
favorable conditions for the development of dynamical instabilities
that lead to the old bulges, and the younger disks that surround the
late-type bulges today could have been easily accreted after the bulge
formation event (see, e.g., Avila-Reese et al.\ 2005).

Another mode of disk evolution that can lead to the formation of
bulges early-on in the history of the Universe is the fragmentation of
gas-rich disks into massive dumps of subgalactic size, whose
subsequent dynamical friction leads to spheroidal bulges embedded in
exponential disks.  In Debattista et al.\ (2006) we present SPH
numerical simulations of live dark matter halos plus a baryonic
stars+gas disk component that we   performed   with GASOLINE (Wadsley,
Stadel \& Quinn 2003). These simulations show that a gas-rich disk can
become violently gravitationally unstable if it cools efficiently by
radiation. The disk fragments into gaseous massive ($M > 10^7$ M$_{\odot}$) clumps. Star formation is enhanced in these clumps, which spiral in towards the center,  dragging with them the associated
stellar clumps. The mergers of clumps produce a bulge component on 
a time scale of   $\sim$10 inner disk orbital times,
i.e., of at most a couple of Gyr.

The ability of the disk to fragment depends on the surface density of the gas component, which in turn scales as the  virial mass in currently
favored galaxy formation models (e.g., Mo, Mao \& White). With a typical $\sim5\%$ disk mass fraction relative to the halo mass, and a typical spin parameter of $\lambda=0.035$, 
a  disk  with $v_{circ} \sim 80$ km/s will fragment into clumps when about 70\% of its  mass is in the gas component (which guarantees a Toomre [1964] parameter Q approximately $<1.4$, a condition necessary for fragmentation of self-gravitating disks, Mayer et al. 2004). 
Given the typical circular velocities of our sample galaxies, it is thus possible that these became clump-unstable at early times, when most of their disk mass was in the gaseous form.

We illustrate the evolution of such a Milky-Way-like progenitor in
Figure~\ref{fig:marcisims}, where we show the face-on projected view
of the gaseous and stellar densities of one of the SPH simulations of
Debattista et al.\ (2006;   see   this paper for further details).
In this specific simulation we sample the dark matter halo, the
stellar disk and the gaseous disk using respectively $10^6$, $2\times
10^5$ and $10^5$ particles, i.e., 50\% of the baryonic mass is
initially in the form of diffuse gas, which represents the partially
ionized hydrogen component of the galaxy. The time interval between
the top and the bottom images in the left panel of the Figure, during
which this numerical model of a Milky-Way-like galaxy   first forms  the massive clumps and then 
transfers them to the center, and thus transitions from
its smooth exponential initial conditions to its final,  bulge+disk 
configuration, is about 1 Gyr.  
The final system is a late-type spiral with an old {\it and} almost
exponential bulge (see the projected density profile shown in the
bottom right panel of Figure~\ref{fig:marcisims}), i.e, a system
quantitatively similar to several of the galaxies that we have
analyzed in this work, and similar to our own Galaxy.

Similar results have been found by Wada \&
Norman (2002), who use multi-phase grid-based simulations which includes supernovae heating in a small sub-volume of a galactic disk, and by  Noguchi (1998), who use different numerical 
techniques as well as different models for the galaxies, and also finds fragmentation of gaseous disks  into massive dumps of subgalactic size that ultimately leads to structures resembling present-day galaxies, with a spheroidal bulge and an exponential disk. 
It is interesting to note that images of high-z galaxies reveal morphologies which match astonishingly well the clumpy density distribution of Figure~\ref{fig:marcisims} (see e.g. Elmegreen et al.
2004). This provides further indirect evidence that this mechanism, 
together with the classical  halo-mergers, monolithic collapse, and, 
possibly, even bar-driven secular evolution of dense  seed disks, is
a possible channel for the formation of old late-type bulges.

\section{Concluding remarks}

The main conclusions of our work are:

{\it (1.)}  In our sample, several late-type bulges have underlying, diffuse
stellar populations with optical/near-infrared blue colors that
strongly support a bulge-after-disk formation scenario.

While satellite accretion onto the disk might be a player in forming bulges at late times, the correlation between disk and bulge scalelengths is possibly more naturally explained by 
bulge formation via disk instabilities (although it is not excluded that these might possibly 
be triggered by interactions). Observations of the high redshift universe show that
the  size-luminosity function of  disk galaxies remains constant up to
redshift of $z=1$ (Lilly et al.\ 1998; Sargent et al.\ 2006), and that a
comparable fraction of bars is observed in disk galaxies at redshift $z=0$ (Eskridge 2002) and
at redshift $z=1$ (Jogee et al.\ 2004; Sheth et al. 2006).  Indeed 
the  disks appear, rather early-on in their lives, to be  massive enough and capable 
of forming bars. It is thus  possibly not surprising that real barred disks are then able   to become    
 unstable to bar dissolution, and/or of fueling the galaxy centers of bulge-building dissipative material, as it is observed in computer models of disk galaxy evolution.

While our sample is too small to draw firm statistical conclusions, our results nonetheless
suggest that a large fraction of the  late-type  galaxies that at $z=0$ host a bulge 
must have formed as pure bulgeless disks, and developed a bulge at later times. 
This population of pure-disks that has evolved into bulged-disks at later epochs adds to the pure exponential disk galaxies which are also observed at $z=0$; these alone account for at
least $\sim30\%$ of the galaxies classified as Sd and later types
(Boeker et al.\ 2003).   The sum implies that there must have been is a large population of bulgeless disk galaxies at high redshift, one that    CDM cosmological  simulations of galaxy formation    at present do not predict, but must  become able to  provide.   
To date,  cosmological simulations have not been able to produce realistic disks with a pure exponential
profile.

{\it (2.)} The bulk of the stellar mass in the remaining  late-type bulges in our sample is in old stellar
populations,  indicating that in some cases  late-type-bulge formation can occur  before the formation of the disks. 

The old late-type bulges in our sample are very similar to the Milky Way Bulge in all respects: Not only most of their stellar mass in rather old stars (for the Milky Way, $\sim$10Gyr: Ortolani et al.  1995; Renzini 1999a, 2002;  Zoccali et al.\ 2003), but also they 
show a similar relationship between disk and bulge scale-lengths, and
have  suffered from some rejuvenation at later times (see van Loon et al.\ 2003 for evidence for young stars in the Galactic Bulge). This  result   extends other recent work on the ages of spiral bulges (Thomas \& Davies 2006) down to the smallest bulge masses. The Milky Way Bulge is thus not  an exceptional late-type bulge, but rather a representative of the old-end-side of the distribution that is observed in the local population. 

{\it (3.)}  It is intriguing that, down to the mass scales of "transition" between nuclear star clusters and bulges, the stellar age of the late-type bulges correlates with the bulge mass, which also however correlates with the galaxy mass. The current data are compatible with  the galaxy mass governing the fraction of stellar mass that ends up in a bulge component, as well as the epoch of bulge  formation. A question that we are currently investigating is whether there are also environmental effects that lead to an early or delayed formation of bulges in the Universe. To achieve this goal we are undertaking the Zurich ENvironmental survey (ZENs),  to study  $>2200$ galaxies that are known members of 185 nearby galaxy groups (selected from the 2PIGG galaxy group catalogue, Eke et al. 2004, which has been constructed starting from the 2dFGRS, Colless et al.\ 2001). The ZENs groups cover a large range of group masses, virial radii, densities, and location relative to the filamentary large-scale-structure in which they are embedded; the groups are also being observed with the GBT at radio wavelengths to map  their HI content. We expect that the ZENs systematic census at redshift zero of bulge properties as a function of group environment, coupled with our ongoing numerical experiments of disk evolution, which we are now extending to explore a range of environmental conditions, will be able to establish the role of the local and large-scale-structure environment in the development of the central regions of disk galaxies.
 
 Finally, an important question that we have left open is  how the properties of the late-type bulges and inner disks that we have presented here relate to the properties of the nuclear star clusters that sit in the centers of these galaxies. We address this issue in Paper 2.
  
\acknowledgments
We thank the anonymous referee and Tim de Zeeuw for constructive comments which have improved the presentation of our results. CS acknowledges support from the Swiss National Science Foundation.
This publication was supported by NASA grant GO-09395.03-A, and
makes use of data products from the Two Micron All
Sky Survey, which is a joint project of the University of
Massachusetts and the Infrared Processing and Analysis
Center/California Institute of Technology, funded by the National
Aeronautics and Space Administration and the National Science
Foundation.

\newpage

\begin{center}{\bf APPENDIX A:  Observations, data reduction and analysis}\end{center}

\bigskip

{\bf A.1 The sample properties, the data and the data reduction}

\smallskip

To illustrate the properties of our small sample relative to the entire late-type disk galaxy population, Figure~\ref{fig:compstructma} shows the comparison between disk scale lengths, rotational velocities, densities, and total magnitudes 
for our nine galaxies (black points) and for the 172 objects of the
ground--based sample of MacArthur et   al.~(2004)   (shaded grey area);
Figure~\ref{fig:struct1}  shows the distribution of  bulge Sersic indices  and  bulge half-light
radii for our galaxies (shaded histograms) and for the
MacArthur et al.\ sample.

The galaxies were observed with the ACS in the F435W ($B$) and F814W ($I$)
filters using the WF camera.  The pixel scale the WF camera is of 0\farcs05
px$^{-1}$, providing   a   total field of view of 3\farcm4$\times$3\farcm4.  We
also observed   our    sample galaxies with the F330W ($U$) filter using the ACS HR
channel, whose pixel scale is 0\farcs027 px$^{-1}$, for a total field of view
of $27''\times 27''$. The $U$ images had too low   a   signal to noise in the
continuum to perform, e.g., isophotal fits (see Appendix~A.2.1); they were thus
not used to investigate the properties of the bulges and inner disks, which
constitute the focus of this paper. We were able however to use the $U$ ACS
data to study the compact sources in the galaxies, including the nuclear star
clusters located in the isophotal centers. We postpone to Paper 2
a detailed discussion   of   the photometrically-distinct sources detected in the
images, and their relation to the host galaxies. 
 
Details   of   the observations, including   the   number of single exposures and exposure
times are listed in Table~\ref{tbl-App1} (for completeness, also for the $U$-band data).
The raw data were processed with the standard ACS pipeline (CALACS) in order
to use the best reference files available at the time of reduction for
flat-field, bias and dark frames, and distortion correction tables. For each
filter, the relative alignment of the individual exposures was checked to be
better than 0.2 pixels.  The individual exposures were then added together
using the IRAF\footnote{IRAF is distributed by NOAO, which is operated by AURA
  Inc., under contract with the National Science Foundation.}  STSDAS task
ACSREJ, in order to obtain a cosmic-ray cleaned image in each filter.  Hot and
cold pixels were partly removed during the combination process, and some
remaining hot pixels were identified and then removed from the final images by
interpolation.

No sky was subtracted from the $U$ images due to the small field of view. Sky
subtraction was however carried out for all galaxies in the $B$ and $I$
images. We first computed the median of the pixel distribution of each image
($S_m$). Contamination by the galaxy and other astronomical sources was
removed by means of a multi-step procedure.  We first rejected all the pixels
with flux greater than $\pm 3\sigma$ from $S_m$. After this $3\sigma$ cutoff,
the background was measured by fitting a Gaussian to the distribution of
values in the remaining pixels.  We compared the results obtained   using   this
method   with measurements of   the sky level in apertures of $5\times5$ pixels located
far from obvious galaxy emission and other sources. We found very good
agreement between the two sky estimates.  For   the   few cases where   it   was
possible, we checked the accuracy of our sky subtraction by comparing the
outer parts of our radial surface brightness profiles with published
ground-based photometry, and found good agreement.

The photometric calibration was done by converting instrumental magnitudes to
the VEGAMAG magnitude systhem by
  applying the phototmetric zeropoints derived from Pavlovsky et al.\ (2005).  We
  used: 25.770 for WFC-F435W images, 25.487 for WFC-F814W, and 22.927 for
  HRC-F330W. All magnitudes were corrected for Galactic extinction following
Schlegel, Finkbeiner \& Davis (1988). Since the ACS filters differ from the
Johnson-Cousins bandpasses, reddening corrections were computed using the
effective transmission curve for each ACS filter created with the task 
CALCBAND in the SYNPHOT package.

\medskip

{\bf A.2 The basic analysis}

\medskip

{\it A.2.1 Isophotal fits}

\smallskip

For each galaxy, surface brightness profiles in $B$ and $I$
were extracted by fitting ellipses to the isophotes using the IRAF
isophote-fitting program ELLIPSE (Jedrzejewski 1987). 

Masks were used in the isophotal fits so as to cover stars, star forming
regions, star clusters, as well as dust patches and lanes. The masks were
obtained by means of an iterative procedure applied to the $B$ images.
Specifically, for each galaxy we performed an isophotal fit to the $B$ image
without constraining the center, the ellipticity, and the position angle of
the ellipses.  The resulting smooth galaxy model was subtracted from the
original data, and the difference-image was normalized by dividing it by the
original image. A mask frame was then obtained from the normalized
difference-image by setting to 1 all the pixels for which the difference
between the model and the data was greater than a given threshold between 
  30\% and 70\%, depending on the galaxy, and setting to zero all other
pixels. This procedure was reiterated until an acceptable mask was obtained,
as judged by visual inspection.  Masks created with this technique are able
to mask the vast majority of contaminating sharp features, but would of course
not affect any smooth component, e.g., a smooth dust distribution, which would
remain undetected.

For each galaxy, the best fit isophotal parameters (center, ellipticity,
position angle) were obtained from the masked   $I$-band   images. The morphological
parameters obtained from the $I$ images were then kept fixed in performing the
final isophotal fits to the $B$ images.

The $B$ and $I$ surface brightness and morphological profiles for the sample
galaxies are presented in Figures~\ref{fig:profilesB} and \ref{fig:profilesI},
respectively. All profiles are available in electronic tabular form
  from the ApJ database (Tables 8 to 28; these also include extended $H$-band 
  NICMOS+2MASS profiles, see below).

\medskip

{\it A.2.2 Bulge-Disk photometric decomposition}

\smallskip

The wide field of view of the ASC-WFC allowed us to extract relatively
extended radial surface brightness profiles up to $\sim$50 arcseconds from the
center, i.e., well into the disk component for the majority of the galaxies.
This allows us to perform more accurate bulge-disk decompositions than
previously obtained for these galaxies by using the WFPC2 data
(Carollo et al.\ 1997; Carollo et al.\ 1998).

The surface brightness profiles where modeled with a bulge plus disk model.
We assumed an exponential radial profile for the disk component
($I_{\rm disk}(r)$), while we used the general S\'ersic
profile (\ser, Sersic 1968) for the bulge component. The analytical expression
used for the S\'ersic profile is given by:

\begin{equation}
I_{\rm bulge}(r)=I_e\, \exp{(-b_n\,[(r/r_e)^{1/n}-1])},
\end{equation}

\noindent
where $n$ is the S\'ersic index, $I_e$ is the bulge effective surface
brightness, $r_e$ is the bulge half-light radius, and $b_n=1.999\,n
-0.372$ (Caon et al.\ 1993). When $n=1$ the S\'ersic profile corresponds
to an exponential law, when $n=4$ it corresponds to a de Vaucouleurs
profile. The two parameters describing the shape of the exponential disk profile are the
disk central surface brightness $I_0$  and the disk scale length $h$.

In order to derive the five parameters describing the galaxy profiles we
iteratively fitted the two-component analytical model to the observed $I$
surface brightness profiles using a non--linear $\chi^2$ minimization based on
the Levenberg--Marquardt method (e.g., Bevington \& Robinson 1992). The $I$ band
profiles were chosen as they are less affected by spurious ripples and
structures induced by residual dust or star formation features.  For each
galaxy, the analytical fits to the observed profile were performed over an
optimized radial range and after convolving the model with the appropriate
instrumental PSF. The structural parameters derived from the analytical $I$-band fits were 
then kept constant when fitting the $B$ and $H$ surface brightness profiles.

The presence of the nuclear
star clusters is a source of uncertainty in
deriving the parameters describing the underling galaxy light (e.g.,
Carollo e al.\ 1998).  As in our  previous analyses (Carollo et al.\ 1998; 
Carollo \& Stiavelli 1998;  Scarlata et al.\ 2004),
we fitted the galaxy model outside the radial region affected by the
nuclear source, and estimated the uncertainties in the derived
parameters by changing the radial range over which the fit was
performed by $\pm$0\farcs1. The range of radii explored for each
individual galaxy is listed in Table~\ref{tbl-2a}.

The PSF-convolved best fit models are overplotted on the observed ACS
surface brightness profiles in Figures~\ref{fig:profilesB},
\ref{fig:profilesI}.  To show the consistency of the derived models
with the NICMOS $H$-band profiles of Seigar et al.\, we present them
overplotted on the $H$ data in Figure \ref{fig:profilesH}. When
possible, we extended the radial coverage of the NICMOS $H$ data by adding
properly matched $H$ surface brightness profiles derived by fitting ellipses to the
public 2MASS data (Jarrett 2004). For NGC~2758, ESO498G5 and
ESO499G37 the 2MASS data had too low S/N and were thus not considered.
For all galaxies, disk and bulge profiles are represented with dotted
and dashed lines, respectively; the total profiles are indicated by a
solid line.  The combined NICMOS+2MASS $H$ surface brightness profiles
are also available in   electronic   tabular form.

In Tables~\ref{tbl-2a} and ~\ref{tbl-2b} we report the best-fit parameters for the bulge
and the disk components of each galaxy.  For the bulges, these are the
bulge half--light radius $r_{e,\rm I}$, the bulge total apparent
magnitude $m_{I,B,H{\rm bulge}}$, the bulge Sersic index $n$.  For the
disk, the listed parameters are the disk scale length $h_{I}$, and the
central surface brightness $\mu_{0;I,B,H}$.

Consistent with other studies (see Kormendy \& Kennicutt 2004, and references therein), 
the Sersic $n$ indices of the nine galaxies are significantly smaller   than   the $n=4$ 
value which identifies a deVaucouleur's profile, and are close to the
$n=1$ value which describes an exponential profile.

\medskip

{\it A.2.3 The color profiles}

\smallskip

The HST/ACS Point Spread Function (PSF) varies with wavelength, field
position, and time, due to a combination of defocus, coma,
astigmatism, and jittering of the telescope
(Sirianni et al.\ 2005).
Since no point sources were available near the nucleus of the observed
galaxies, we created the appropriate $B$ and $I$ PSFs for each galaxy
using the TinyTim software (Krist \& Hook 2001). We first computed
geometrically distorted PSFs in the observed position of the center of
each galaxy.  The distorted PSF was then corrected for geometric
distortion and rotated using DRIZZLE with the same parameters we used
to combine the galaxy exposures. With this procedure, the spikes of
the   synthetic   PSF are aligned with those of the science images.

We used the PSFs to convolve the analytical models to the surface brightness
profiles before performing the fits (as this procedure is more stable to PSF
uncertainties than deconvolving the data). Furthermore, for the purpose of
deriving the $B-I$ color profiles, the $B$ and $I$ final frames were matched
to the same PSF by convolving each band with the PSF of the other band. We
then derived the $B$ and $I$ surface brightness profiles on these
PSF-convolved images, by performing the same isophotal analysis described in
the previous section, and the $B-I$ color profiles by subtracting the
PSF-$B$-convolved $I$ surface brightness profiles from the PSF-$I$-convolved
$B$ profiles.

Figure~\ref{fig:colorprofiles}  (discussed in the main text)  show the $B-I$ color profiles as a
function of distance to the center normalized to the bulge effective
radius $r_e$.  In order to increase the signal--to--noise ratio of the
color profiles, we binned the surface brightness profiles of the
PSF--crossconvolved images in steps of 1 $r_e$.

In Figure~\ref{fig:colorprofiles}, the innermost points of the color profiles
corresponds to the average bulge color as derived from the photometric
decomposition discussed above.  The observed color profiles have different
properties from galaxy to galaxy.  However, a general feature of the color
profiles is the similarity of the color of the bulge with respect to the
colors of the inner disk, which we discuss in Section 3.3.

\medskip

{\bf A.3 The effects of dust}

\smallskip

Although scattering from dust is expected to be small in galaxies with a
relatively low inclination as in our sample (as the probabilities that photons
are scattered into or out the line of sight are similar; Byun et al.  1994),
there is no doubt that extinction by dust can alter the colors and color
gradients that we are using as diagnostics of stellar population properties.
However, in contrast with ground-based studies of color images, in which sharp
dust lanes and patches would be smeared out at the typical $\sim1"$ seeing,
our analysis, conducted at the resolution of the ACS camera, allows us to
resolve these sharp features, which are masked out in our analysis.  Of
course, a dust component that would appear as smooth and diffuse at the ACS
resolution would remain undetected. If such a component is present, its
spatial distribution and effective optical depth remain unconstrained.
We therefore investigated its impact  on our main conclusions, for two
different  dust models.

In   the   first place we adopted   a   simple foreground screen model to quantify the
effects of dust extinction on colors, using the extinction curve from 
Cardelli et al.\ (1989). Although the approximation of a foreground screen is clearly
unrealistic, such a model is used in most of the literature, and it is thus
convenient to consider it here as well, for a straightforward comparison with
other works. We show, in the Figures reporting our analysis, the
foreground-screen reddening vector for an extinction $A_V$ in the $V$ band of
0.5 magnitudes.

\begin{figure}
\begin{center}
\includegraphics[scale=0.6]{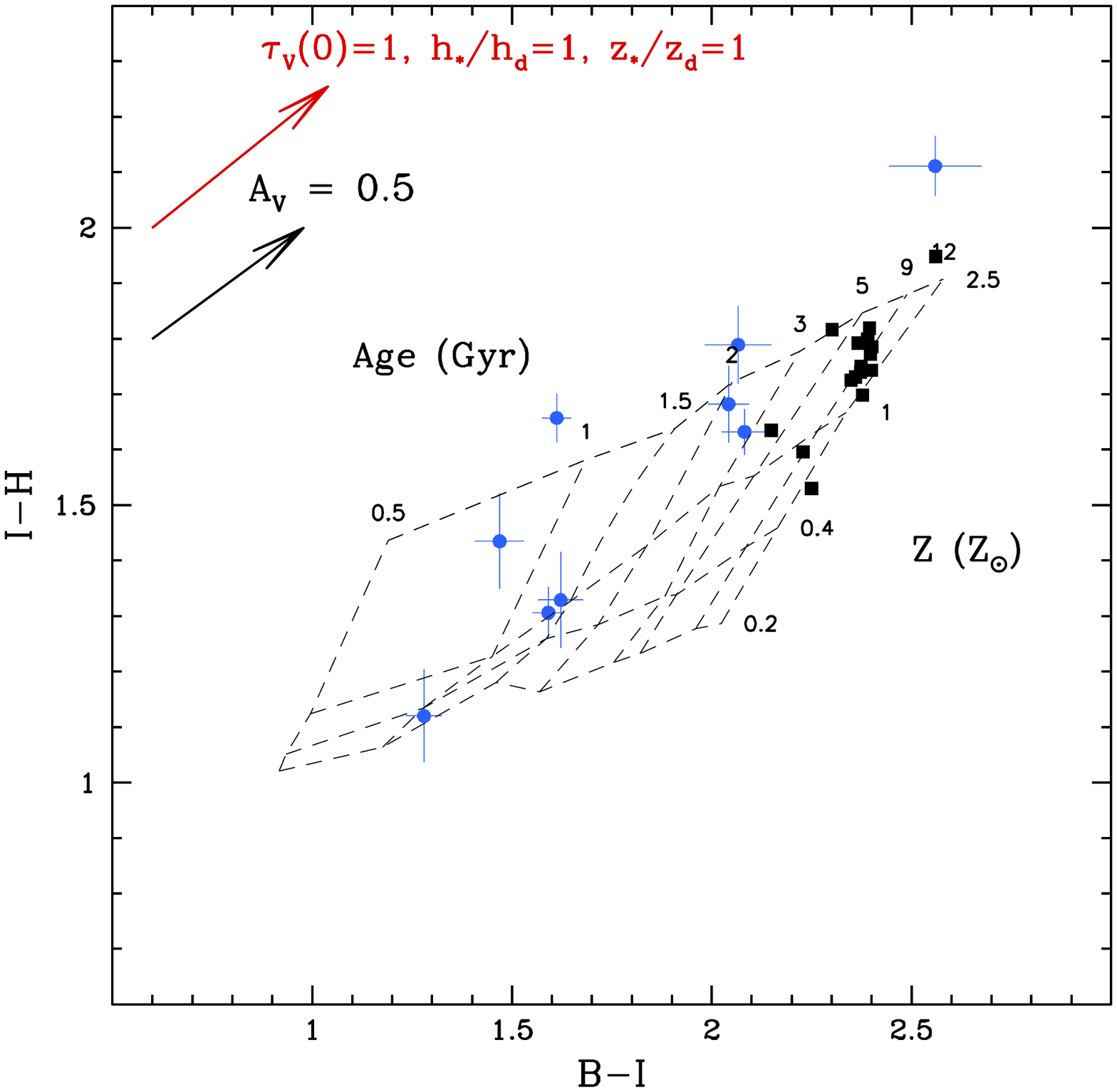}
\caption{The $I-H$ versus $B-I$ color-color diagram  for the nine late-type bulges of
  our sample (blue filled circles). The error bars include the effects of
  varying the radial range for the analytical fits to the surface brightness
  profiles, and the errors in the sky subtraction.  For comparison, the black
  squares represent  massive, early-type bulges of
  Peletier et al.\ (1999). Overplotted is the thoretical grid produced by using
  the BC01 Simple Stellar Populations models.  The almost-horizontal dashed lines are isometallicity tracks with
  metallicity ranging from $Z=0.2Z_{\odot}$ to $Z=2.5Z_{\odot}$.  The
  almost-vertical dashed lines are isochrones of age ranging from 0.5 to 12
    Gyr .  The reddening vectors for a foreground screen (black line) and for
  the more realistic dust model discussed in the appendix (red
  line) are also shown in the upper-left quadrant of the plot
  ($A_V=0.5$ mag).
\label{fig:MIvsIH}
}
\end{center}
\end{figure}

\begin{figure}
\begin{center}
  \includegraphics[scale=0.25]{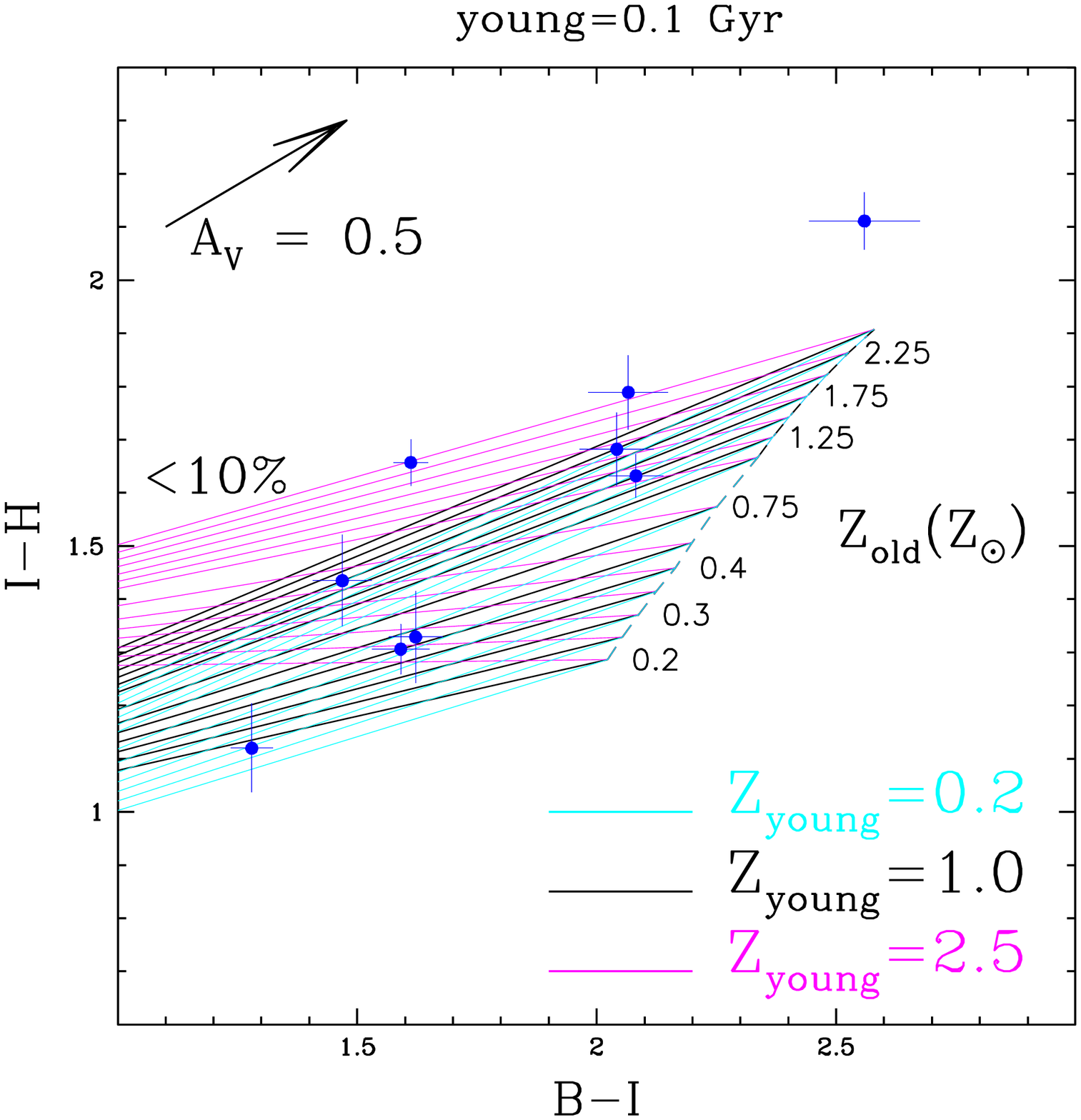}
  \includegraphics[scale=0.25]{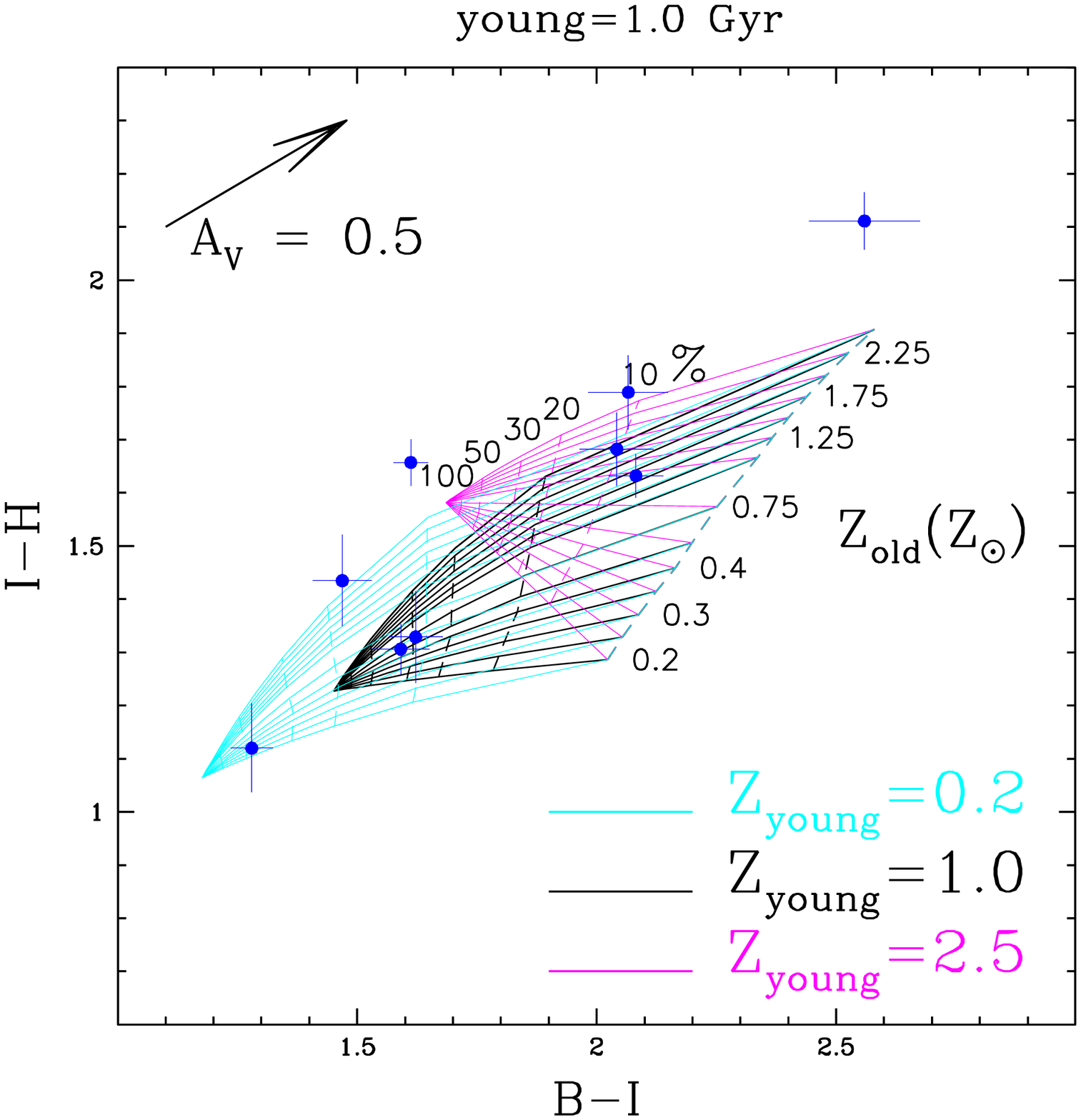}
  \includegraphics[scale=0.25]{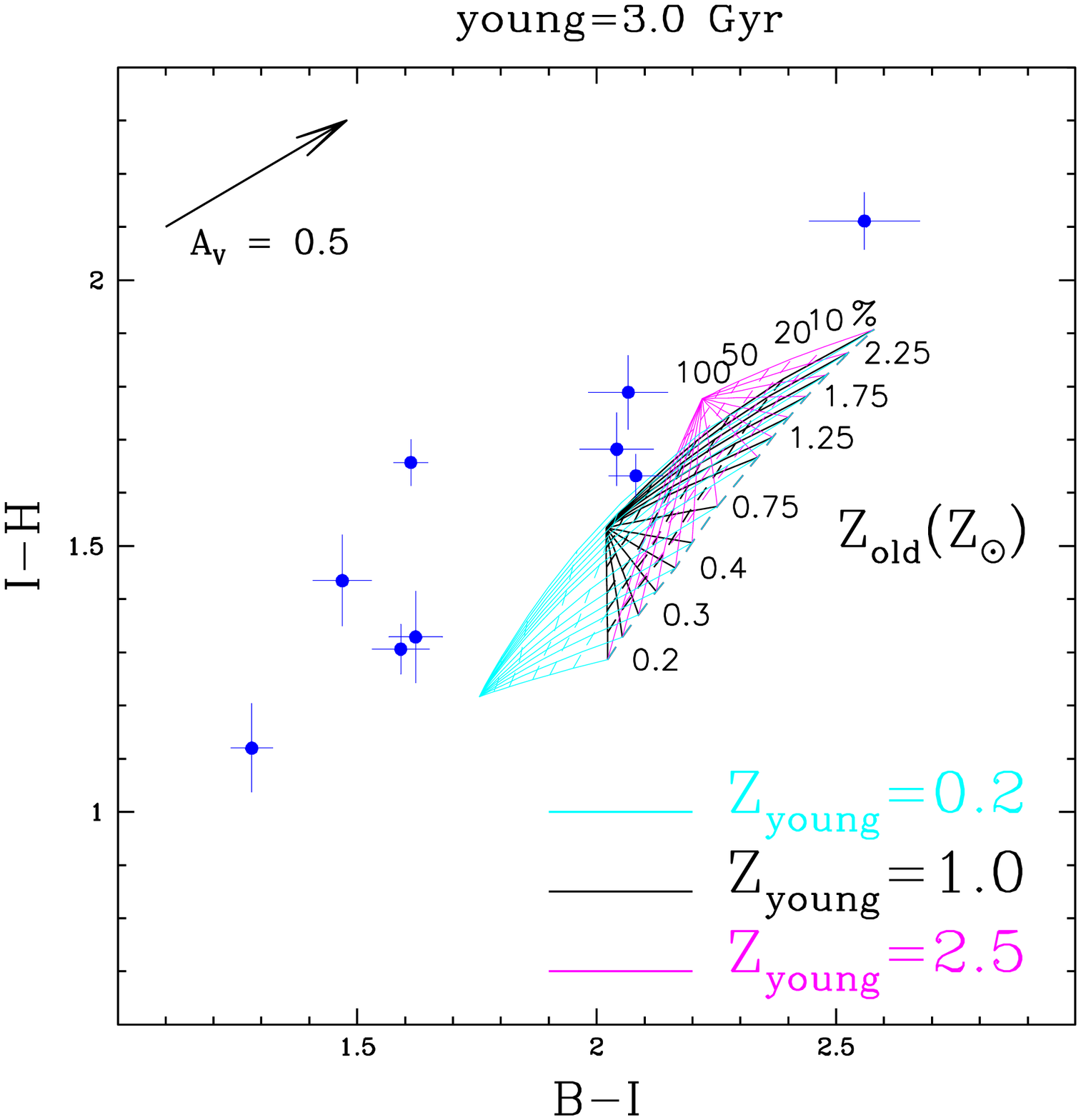}
\caption{The $I-H$ versus $B-I$ color-color diagram with overplotted the grid
    of colors derived for the BC01-based Composite Stellar Population models discussed in the text. In each panel,  the age of the
    young component is kept fixed (top label of each plot), and only its
    fraction in mass and its metallicity are allowed to vary.  For clarity, we
    only plotted tracks corresponding to $0.2, 1.0,$ and 2.5   $Z_{\odot}$  for
    the young component (cyan, black and magenta tracks, respectively). The
     metallicity of the  12~Gyr  old stellar
    component is shown to the right of the tracks. The blue symbols   represent    the late-type bulges in our sample.
\label{fig:COMPOSIT}
}
\end{center}
\end{figure}

\begin{figure}
\begin{center}
  \includegraphics[scale=0.6]{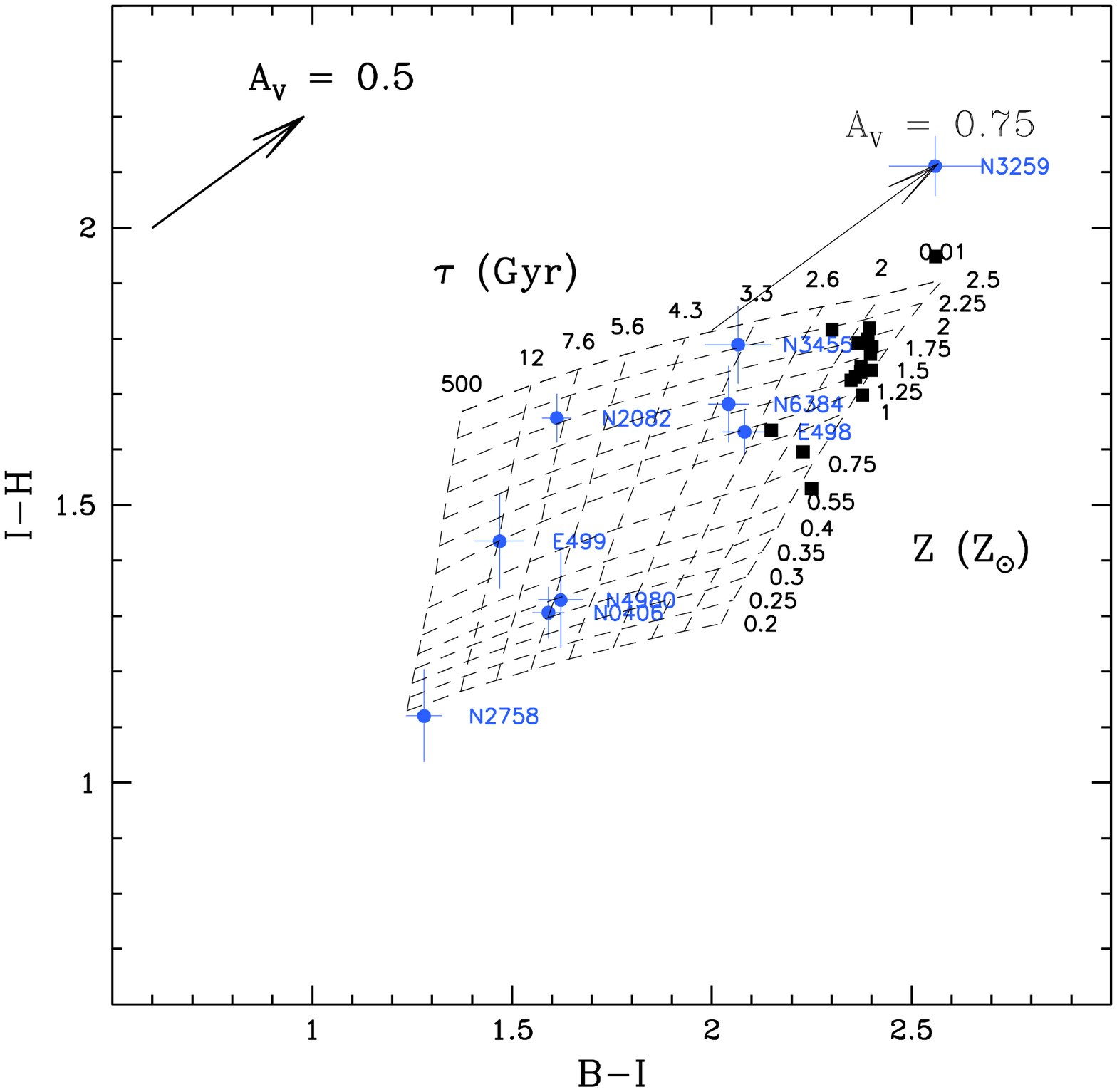}
\caption{The $I-H$ versus $B-I$ color-color
  diagram with overplotted the grid of colors derived for the BC01-based Exponential Star Formation History  models discussed in the text.
  The e-folding time of star formation ($\tau$) is used to label the
  almost-vertical luminosity-weighted isochrones (i.e., tracks along which the
  average luminosity-weighted stellar age, and thus  $\tau$ remain  constant). A very large $\tau$ approximates a constant star formation rate; a very short $\tau$ approximates an instantaneous burst of star formation.
  All  models are assumed to start forming stars   12~Gyr   ago. The metallicity of
  the different models varies from 0.2 to 2.5 $Z_{\odot}$, as shown on the
  right side of the plot. The grey arrow shows the minimum amount of dust (A$_V$=0.75) required to bring NGC~3259 to overlap with the model grid. Other symbols are as in  Figure~\ref{fig:MIvsIH}.
\label{fig:ESFH}
}
\end{center}
\end{figure}

\begin{figure}
\begin{center}
\includegraphics[scale=0.7]{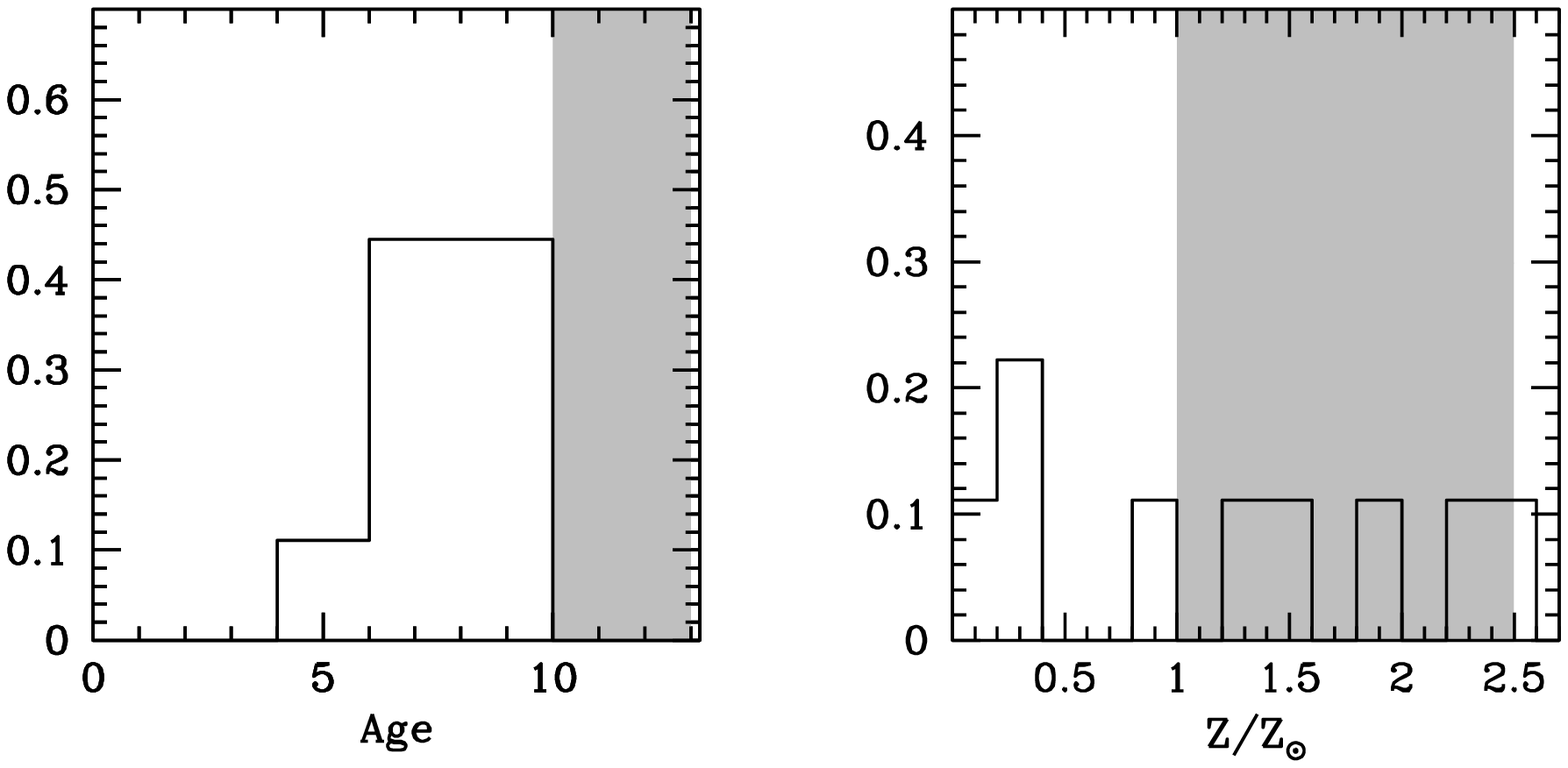}
\caption{Fraction of bulges with a given average stellar age
  ({\it left panel}) and metallicity ({\it right panel}), as derived from the
  BC01-based  ESFH models. The histograms represent the late-type bulges of our sample;
  the shaded areas show the range of ages and metallicities valid for the
  early-type bulges of Peletier et al.\ (1999).
\label{fig:comppel}
}
\end{center}
\end{figure}

\begin{figure}
\begin{center}
\includegraphics[scale=0.4]{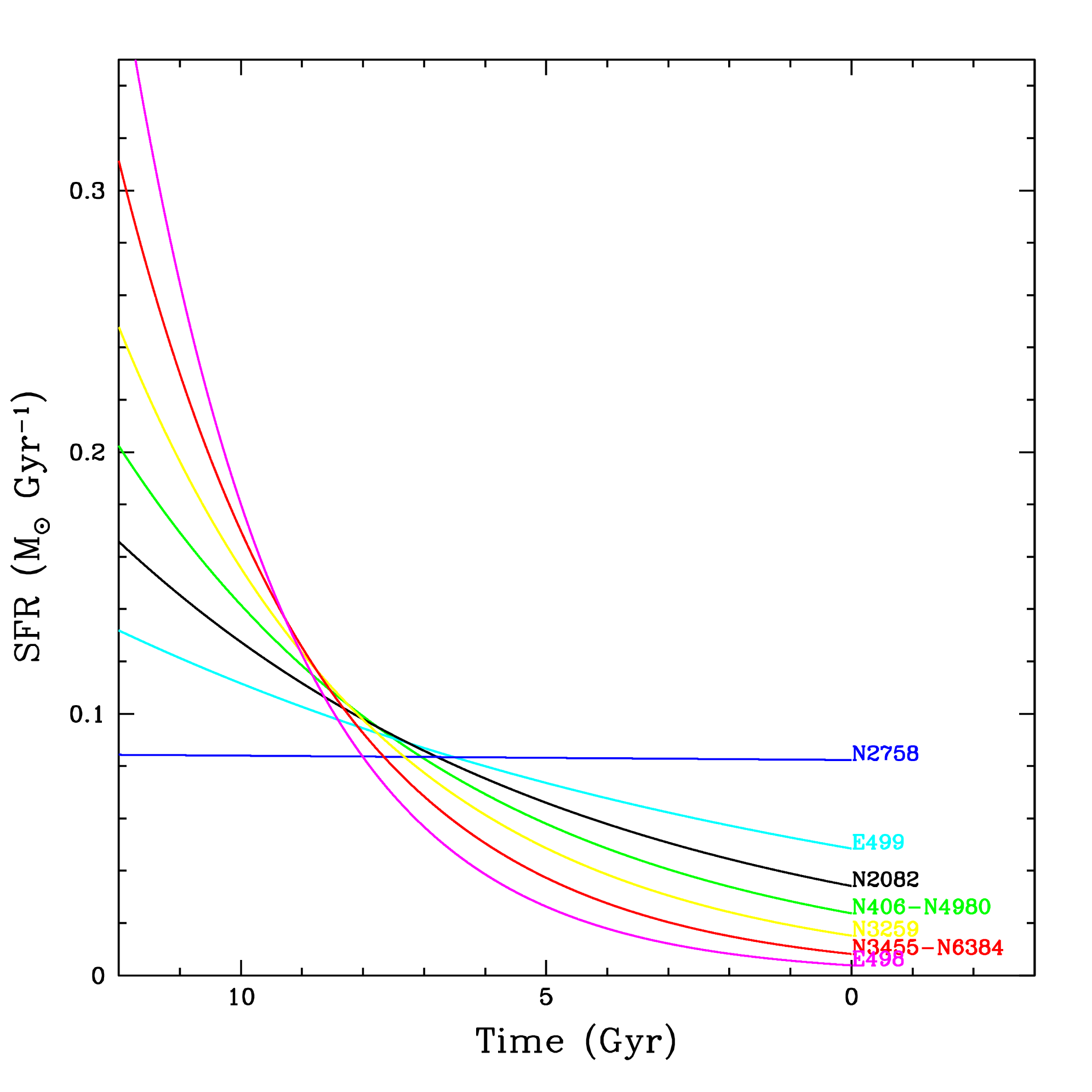}
\includegraphics[scale=0.4]{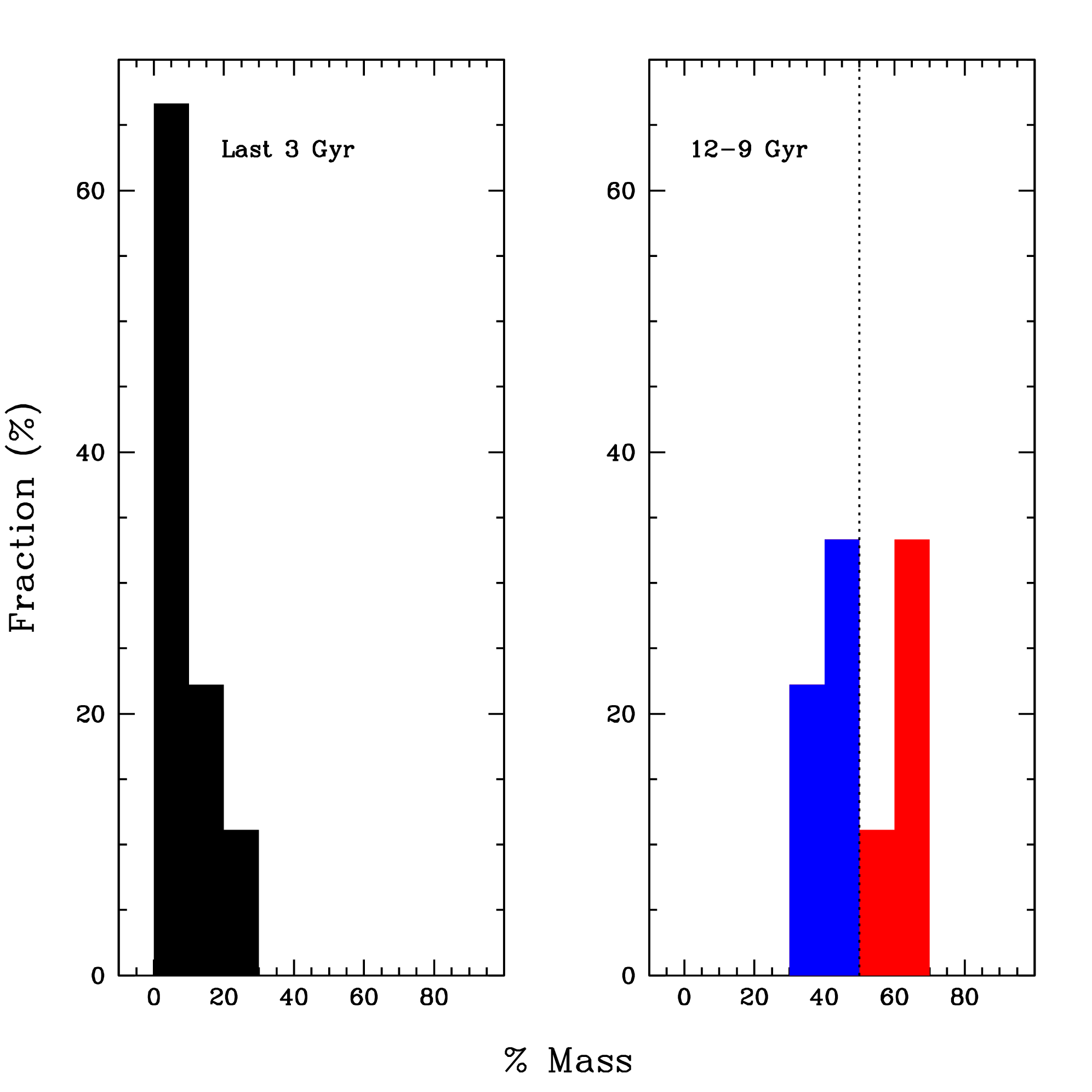}
\caption{{\it Left panel}:
  The best-fit star formation histories obtained from the BC01-based ESFH models for the
  galaxies of our sample. On the x-axis, zero indicates the present time. For
  all galaxies, star formation is assumed to have started   12~Gyr   ago.  {\it
   Right panel}: The vertical axis indicates the fraction of galaxies and the
  horizontal axis the fraction of stellar mass. The histograms show the
  distribution of the stellar mass fraction in the ESFH of the nine
  galaxies that formed in the last   3~Gyr   (left side of figure, shaded histogram) 
  and in the first   3~Gyr   (right side of figure). In the latter, the histogram is shown in red for the fraction of galaxies that formed more than 50\% of their stellar mass more than   9~Gyr   ago, and in blue for the remaining galaxies.
\label{fig:sfhhistmass}
}
\end{center}
\end{figure}

\begin{figure}
\begin{center}
  \includegraphics[scale=0.35]{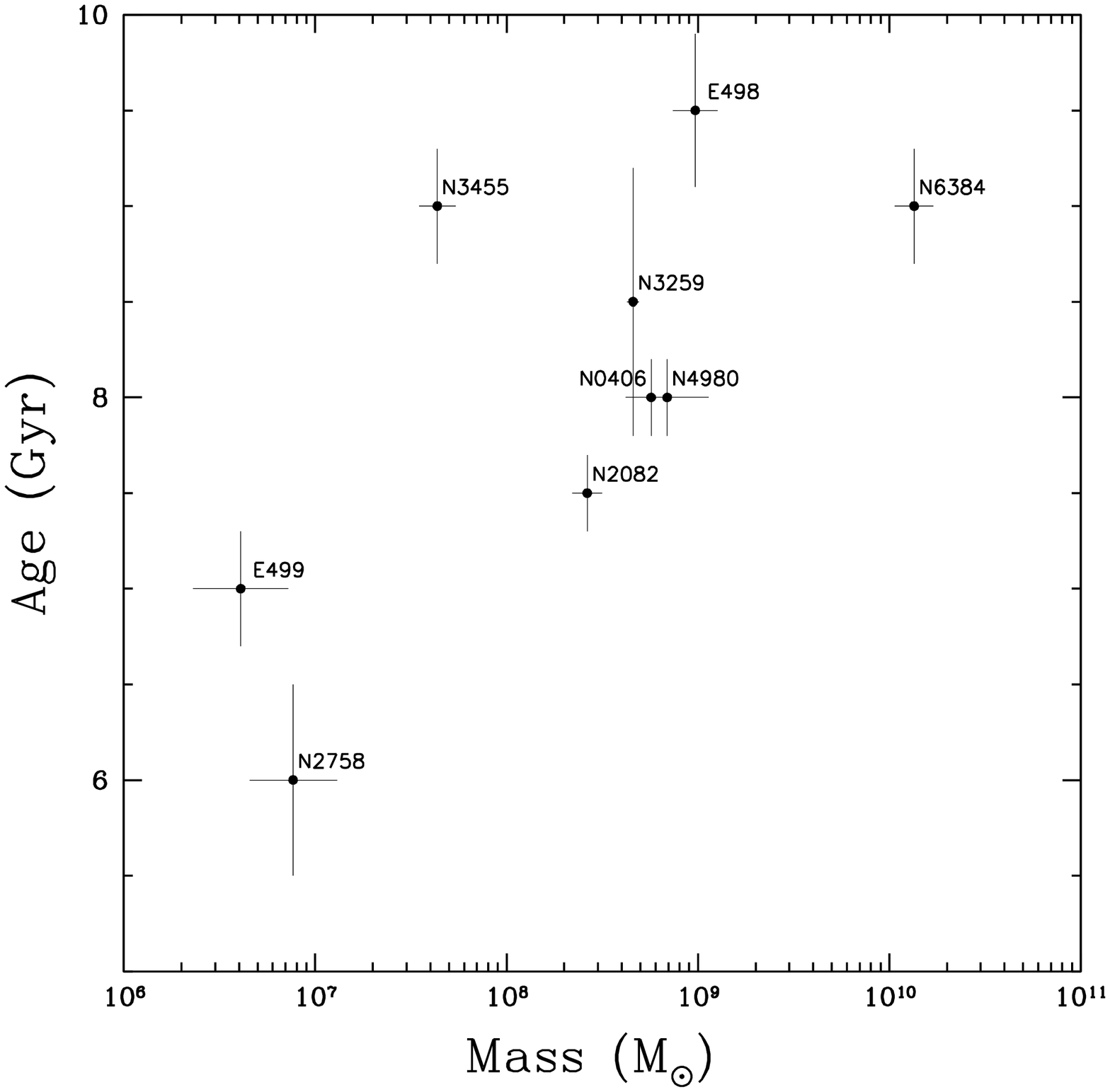}
  \includegraphics[scale=0.35]{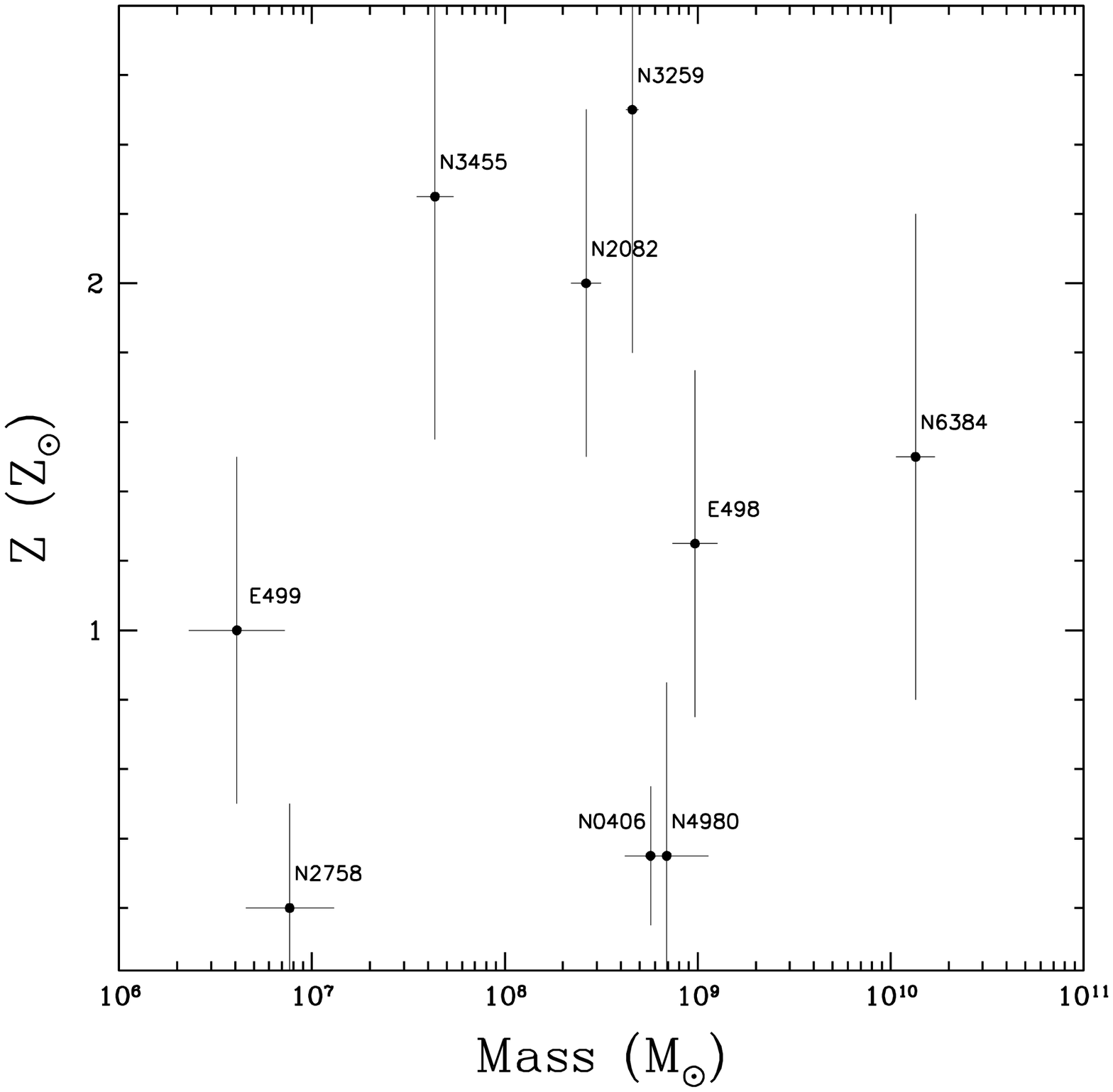}
 \caption{{\it Left panel:} Luminosity-weighted stellar age,  as derived for the late-type bulges of our sample from   the BC01-based ESFH models, versus stellar mass, as measured from the $H$ band
   luminosity of the bulges.  {\it Right panel:} The metallicity of the models that provide the best fits to the colors of the   late-type bulges versus stellar mass.
\label{fig:spmass}
}
\end{center}
\end{figure}

\begin{figure}
\begin{center}
  \includegraphics[scale=0.35]{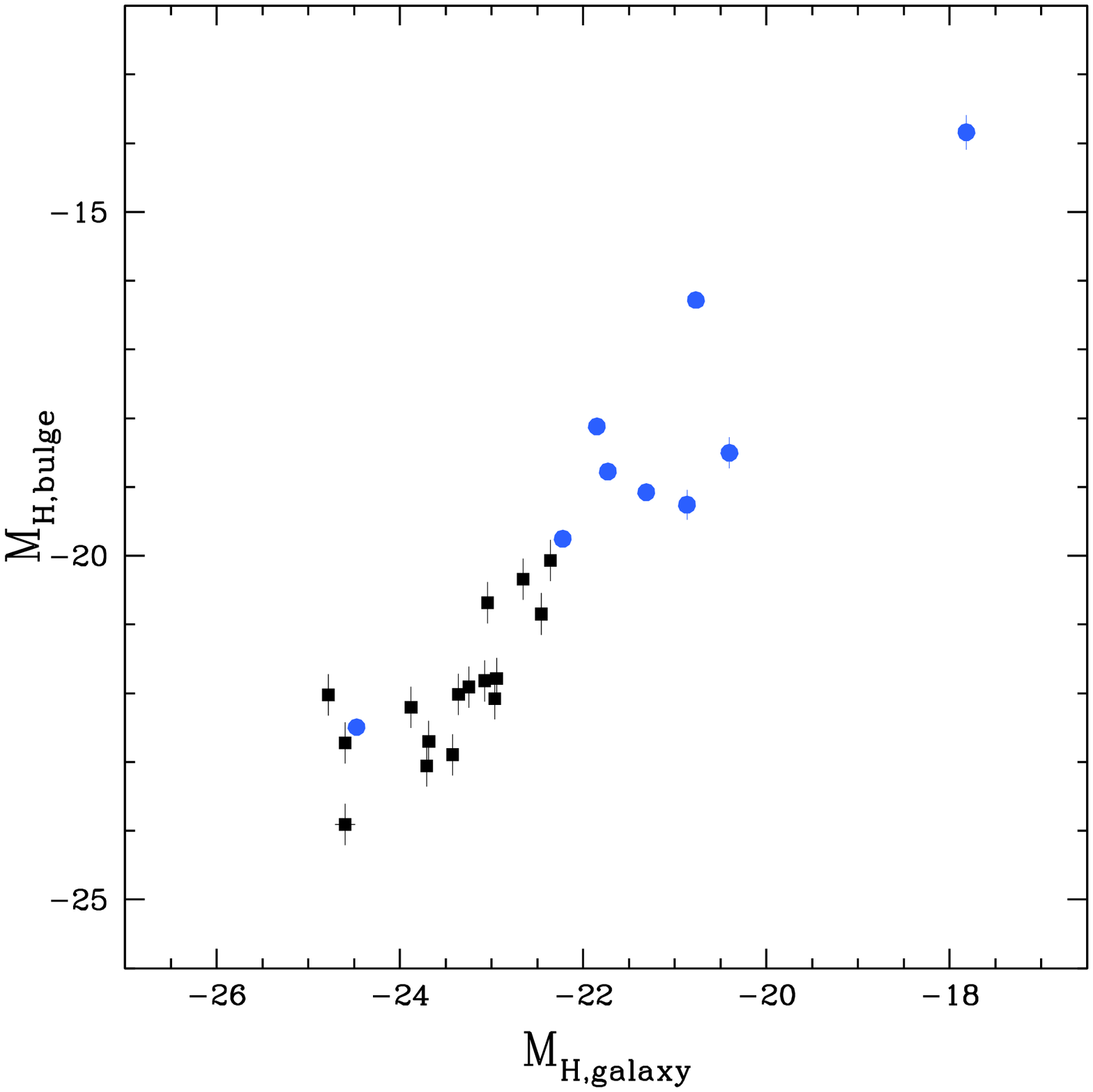}
  \includegraphics[scale=0.35]{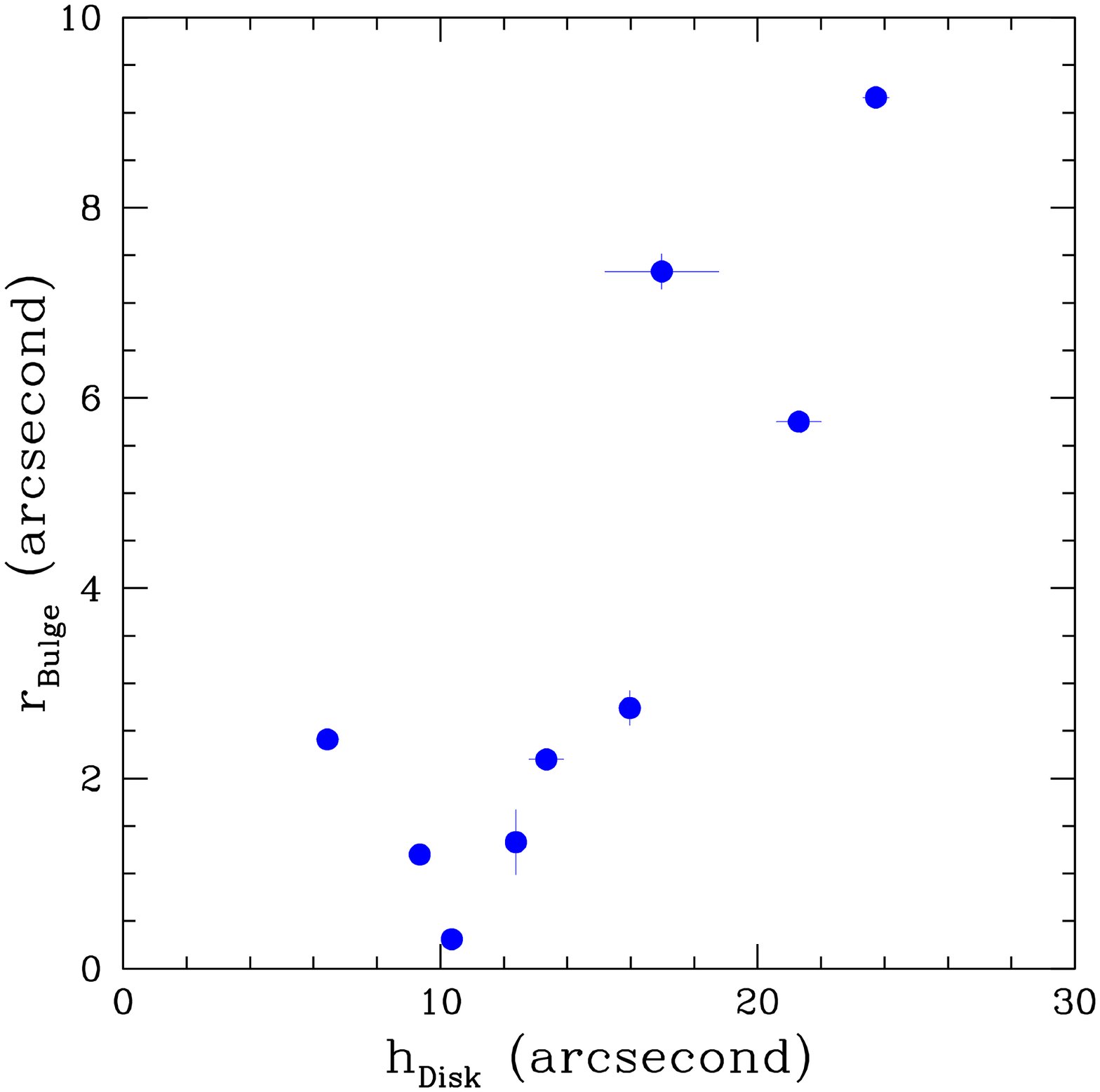}
  \caption{{\it Left panel:} The correlation between bulge and host galaxy $H$-band absolute magnitude (blue: our late-type bulges; black: the early-type bulges of Peletier et al.). {\it Right panel:} The correlation between bulge half light radius and disk scalelength for the galaxies in our sample.
  \label{fig:bulgetot}
}
\end{center}
\end{figure}

\begin{figure}
\begin{center}
 \includegraphics[scale=0.8]{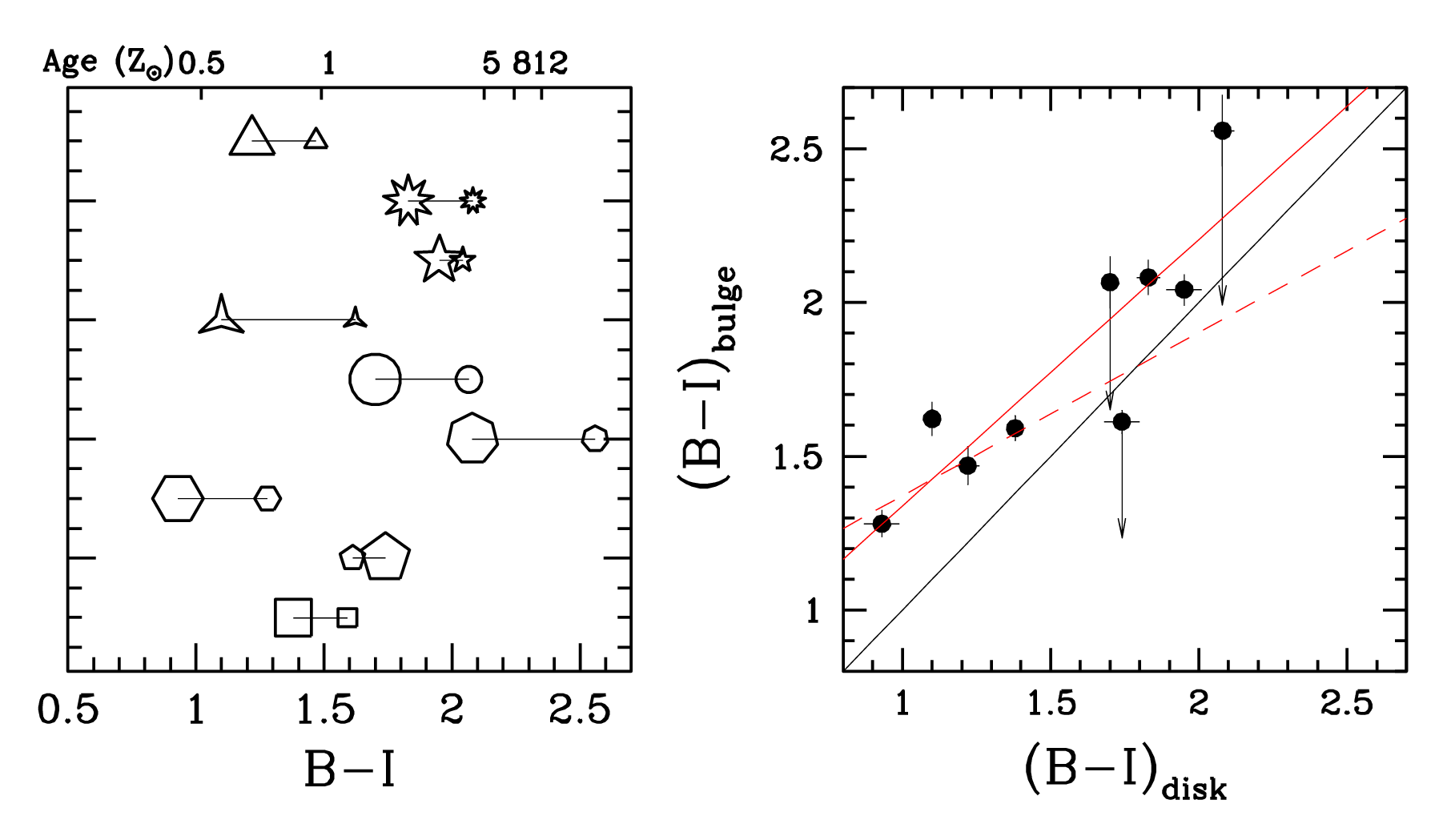}
\caption{{\it Left panel}: $B-I$ color of bulges ({\it small symbols})
  together with the colors of the inner disk, as measured at 5 bulge $r_e$'s
  ({\it large symbols}). Only as a benchmark to guide the eyes, in the top
  horizontal axis we indicate the stellar age corresponding to the associated
  $B-I$ color for a BC01 Single Stellar Population model of solar metallicity. {\it Right panel}: Bulge $B-I$
  as a function of inner-disk $B-I$ (as measured at 5 $r_e$). The red solid line is
  the best fit  to the observed points [$(B-I)_{\rm bulge}=(0.9
  \pm 0.2)(B-I)_{\rm disk}+0.5\pm 0.3$]; the dashed red line represents
  the best fit derived from   our estimates of   the de-reddened colors (see text for details).
\label{fig:bulgeinnerdisk}
}
\end{center}
\end{figure}

\begin{figure}
\begin{center}
\includegraphics[scale=0.35]{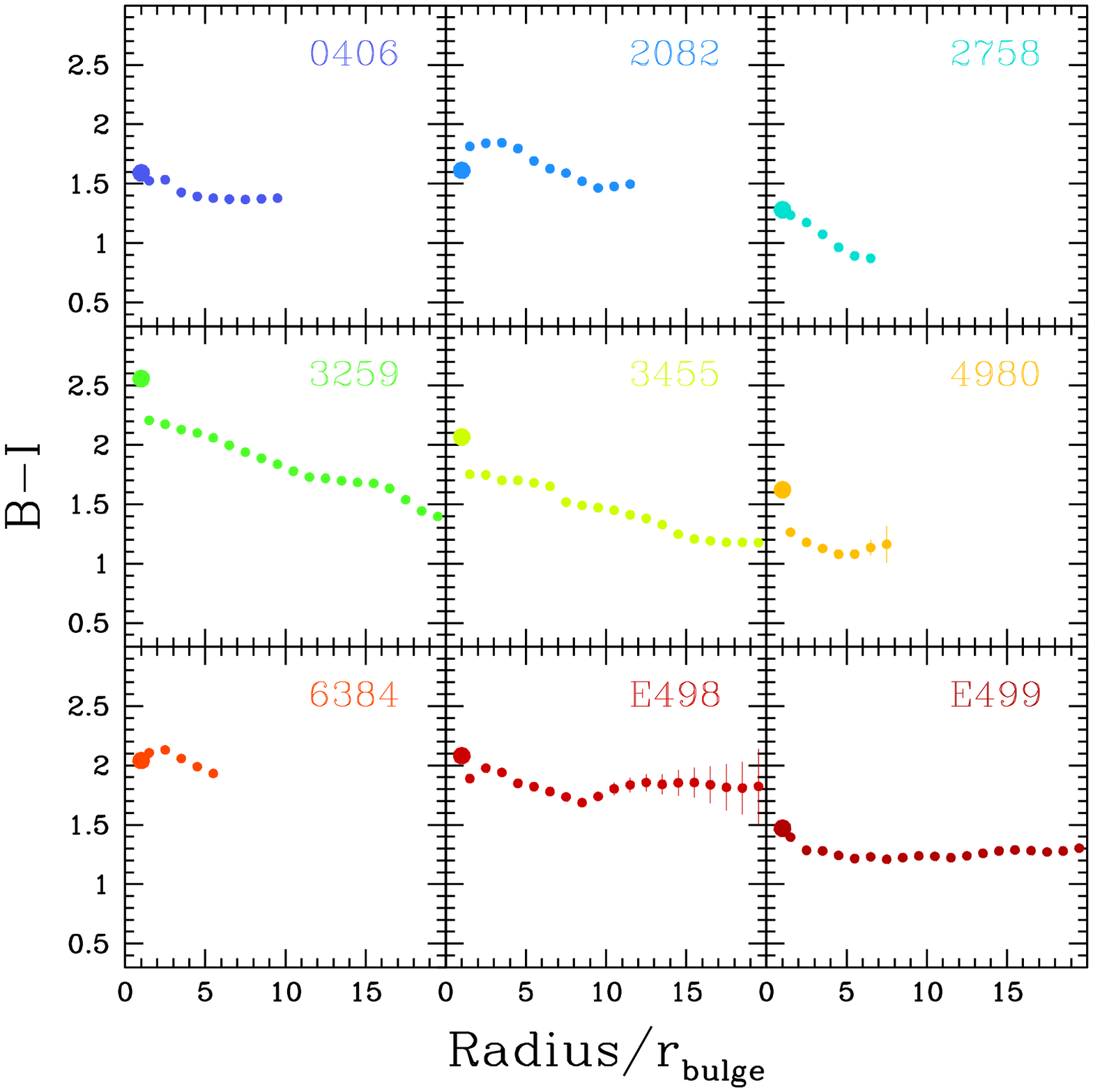}
\includegraphics[scale=0.35]{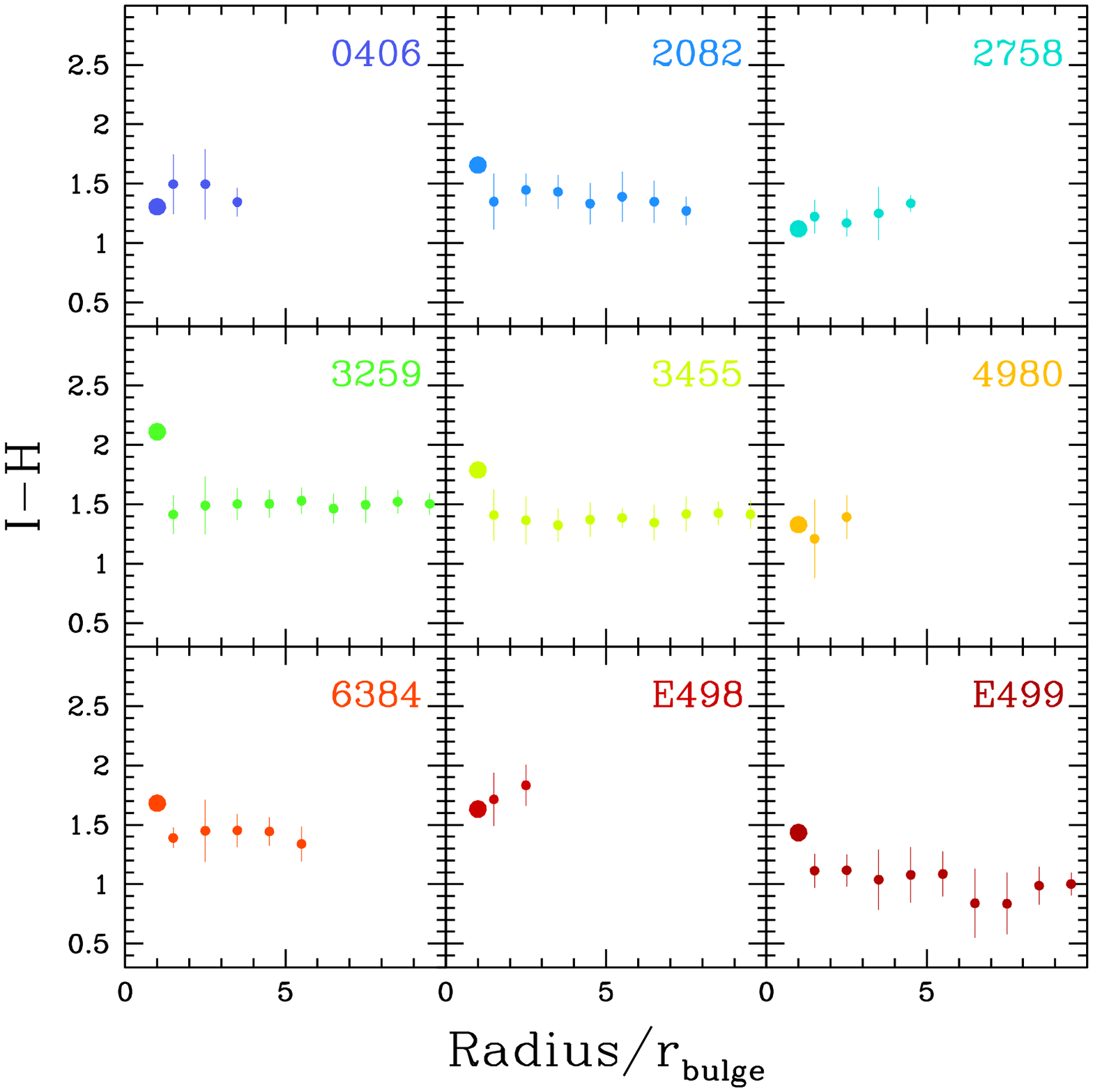}
\caption{The $B-I$ and $I-H$ radial profiles for the 
    sample galaxies as a function of distance from the center, in units of the $I$-band
    bulge half-light radius $r_e$. The first (large symbol) points, plotted at
    1$r_e$, indicate the total bulge colors derived adopting the bulge
    components of the analytical fits to the surface brightness profiles (performed keeping in 
    all bands the structural parameters fixed to the best analytical fits to the $I$-band profiles). 
     The other points at $r/r_e>1$ have been averaged within bins of 1$r_e$. The
    nominal inner-disk colors that we use in our discussion are the color values
    measured at 5$r_e$'s.
\label{fig:colorprofiles}
}
\end{center}
\end{figure}

\begin{figure}
\begin{center}
  \includegraphics[scale=0.6]{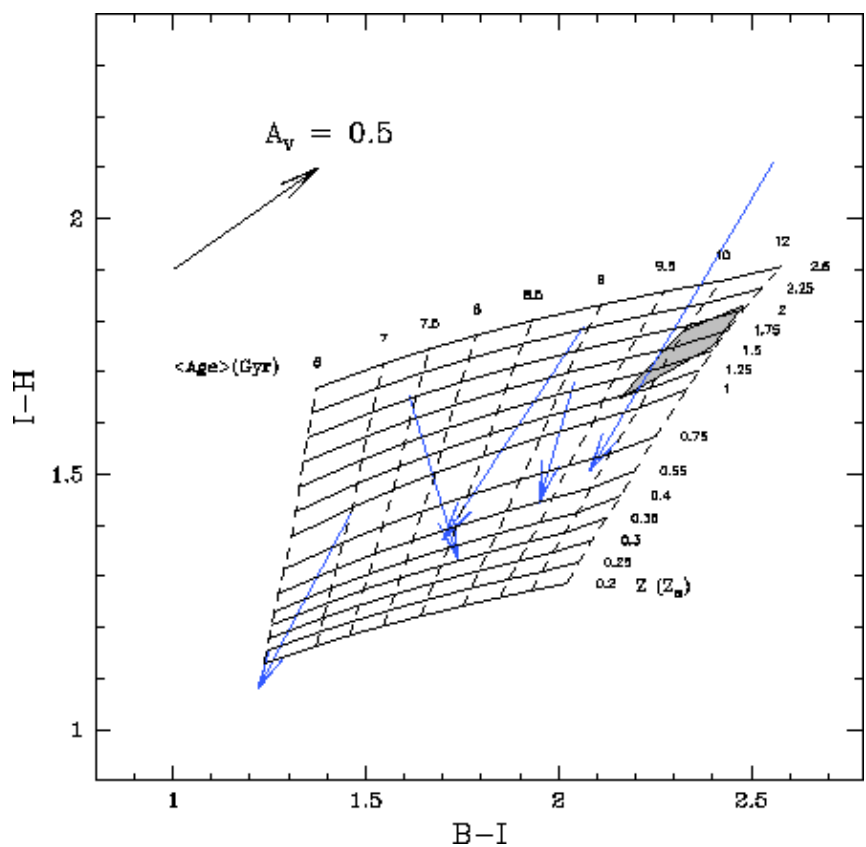}
 \caption{ Comparison on the $I-H$ versus $B-I$ color-color diagram between
   bulges ({\it start-point of arrows}) and inner disks ({\it
     end-point of arrows}), for the five late-type galaxies of our
   sample for which the $H-$band surface brightness profiles extend
   far enough to allow us to measure the inner-disk $I-H$ color at 5
   bulge $r_e$'s, consistently with the $B-I$ measurements.  The grey
   area represents the average bulge-disk gradient measured in
   early-type spirals, which we derived from the data published in
   Peletier \& Balcells (1996). For comparison, the grid of BC01-based ESFH
   models is also overplotted.
\label{fig:bulgeinnerdisk2}
}
\end{center}
\end{figure}

\begin{figure}
\begin{center}
\includegraphics[scale=0.6]{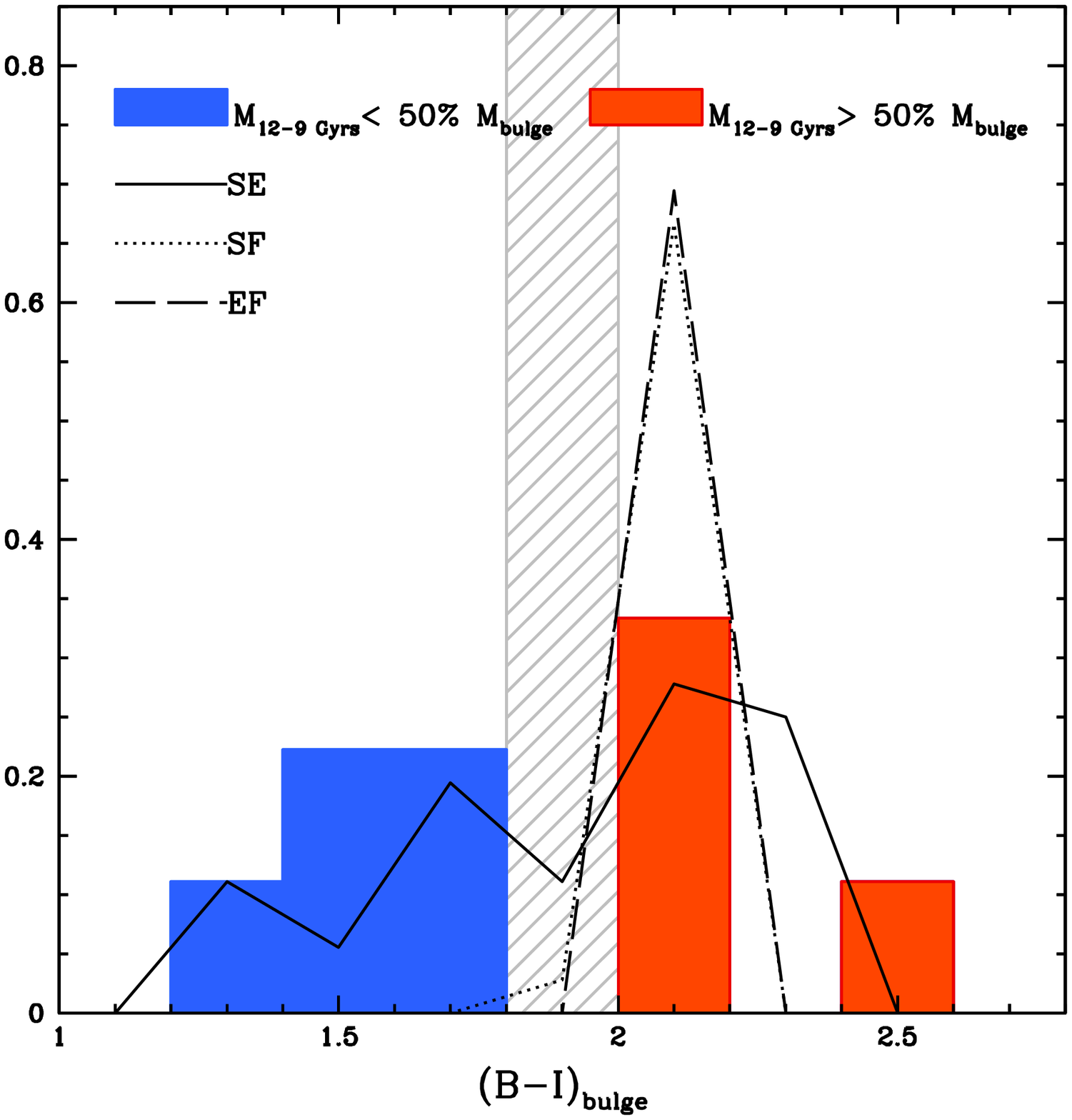}
\caption{ 
The $B-I$ color distribution for the late-type bulges in our sample that formed
  more than 50\% of their mass in the first 3 Gyrs of their ESFHs (red
    histogram), and of those that formed less than 50\% in the same
  period (blue histogram). Solid, dotted and dashed
  lines represent the predictions of Bouwens et al.\ (1999)  for the so-called
  Secular Evolution (SE), Simultaneous Formation  (SF) and
  Early Formation (EF) bulge models, respectively (see
  text for details). The grey shaded area highlights the $B-I$ color (with a
  $\pm 0.1$ mag uncertainty)
  below which the SF and EF models cannot explain the observed colors of
  bulges.
\label{fig:bouwens}
}
\end{center}
\end{figure}

\begin{figure}
\begin{center}
\includegraphics[scale=0.45]{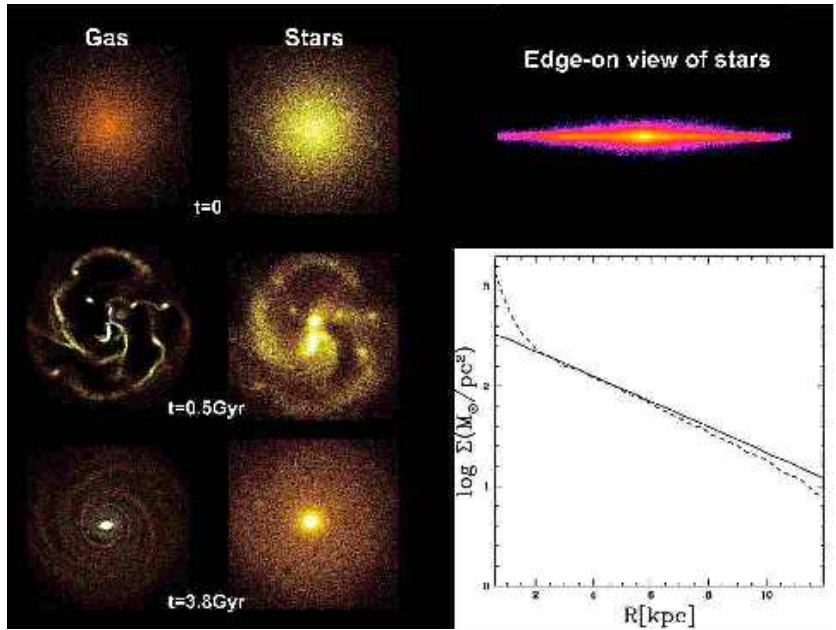}
\caption{ {\it Left panel}: An SPH simulation with initial conditions
  suited to represent a Milky-Way-like galaxy, with an initial gas
  fraction of 50\% in the disk. The simulation includes radiative
  cooling. The left images show, from top to bottom, the gas density
  after zero, 0.5 and 3.8 Gyr from the start of the simulation. The
  right images show the stellar density at the same snapshots in time.
  After less than one Gyr, the sinking into the galaxy centers of the
  clumps seen in the central panels transforms the initially perfectly
  exponential disk into a disk with a three-dimensional bulge similar
  to the Milky Way bulge.  {\it Right panels}: The upper figure shows
  an edge-on view of the stellar density at the end of the simulation;
  the lower figure shows the projected stellar density profile, in which the
  solid line is used for the initial profile and the dashed line for the
  profile after 3.8 Gyr as viewed at an inclination of 60 degrees. The
  final profile is well represented by the sum of an exponential disk
  and a low (n$\sim 0.5$) Sersic-index bulge.
\label{fig:marcisims}
}
\end{center}
\end{figure}

\begin{figure}
\begin{center}
\includegraphics[scale=0.6]{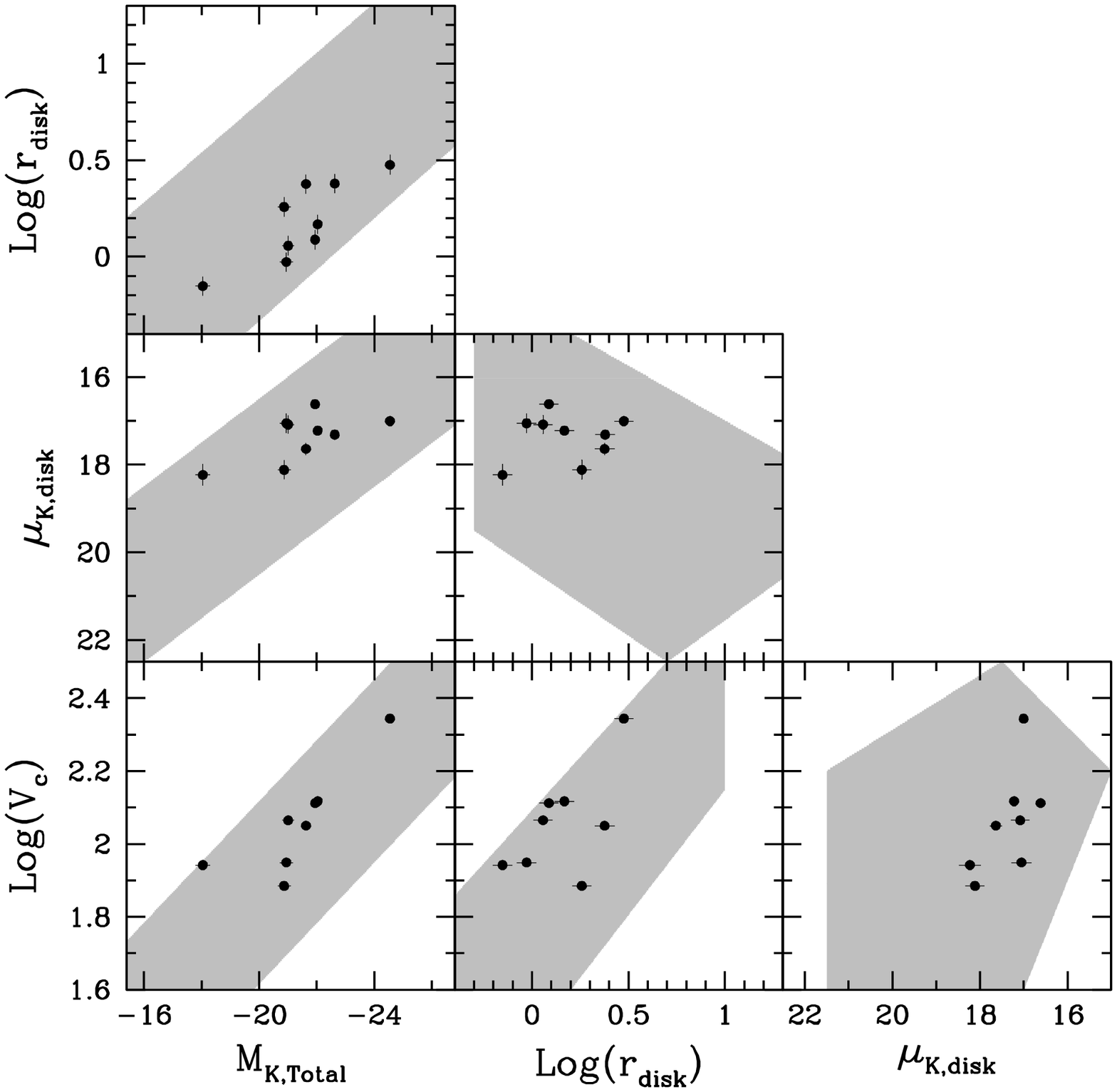}
\caption{A comparison of the structural parameters for the nine galaxies of our
  sample ({\it black points}) with the whole parameter space covered by the
  late--type disk galaxy population ({\it shaded areas}; information extracted from
  MacArthur et al.\ 2004). The plotted quantities are: the bulge Sersic index $n$, 
  the logarithm of
  the circular velocity $v_c$ (in km/s$^{-1}$), the central surface brightness
  of the disk $\mu_{K,disk}$ (in mag/arcsec$^2$) and the logarithm of the disk
  scalelength (in kpc). For our galaxies, we converted the available $H$-band
  disk central surface   brightness   $\mu_{H,disk}$ into the $K$ band value using the
average $H-K$ color as a function of Hubble Type presented in Table~1
of Macarthur et al.\ (2004).
\label{fig:compstructma}
}
\end{center}
\end{figure}

\begin{figure}
\begin{center}
\includegraphics[scale=0.4]{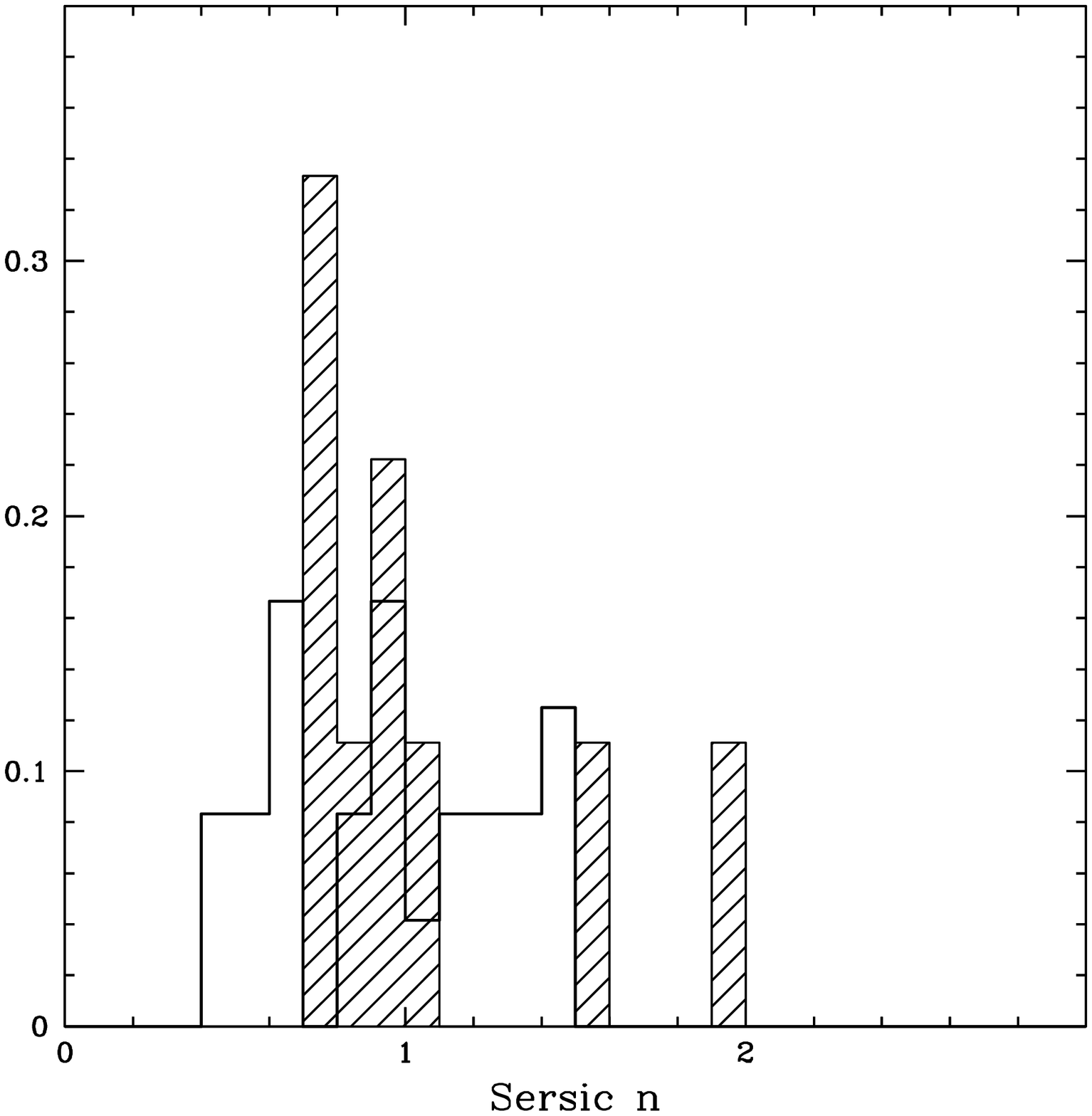}
\includegraphics[scale=0.4]{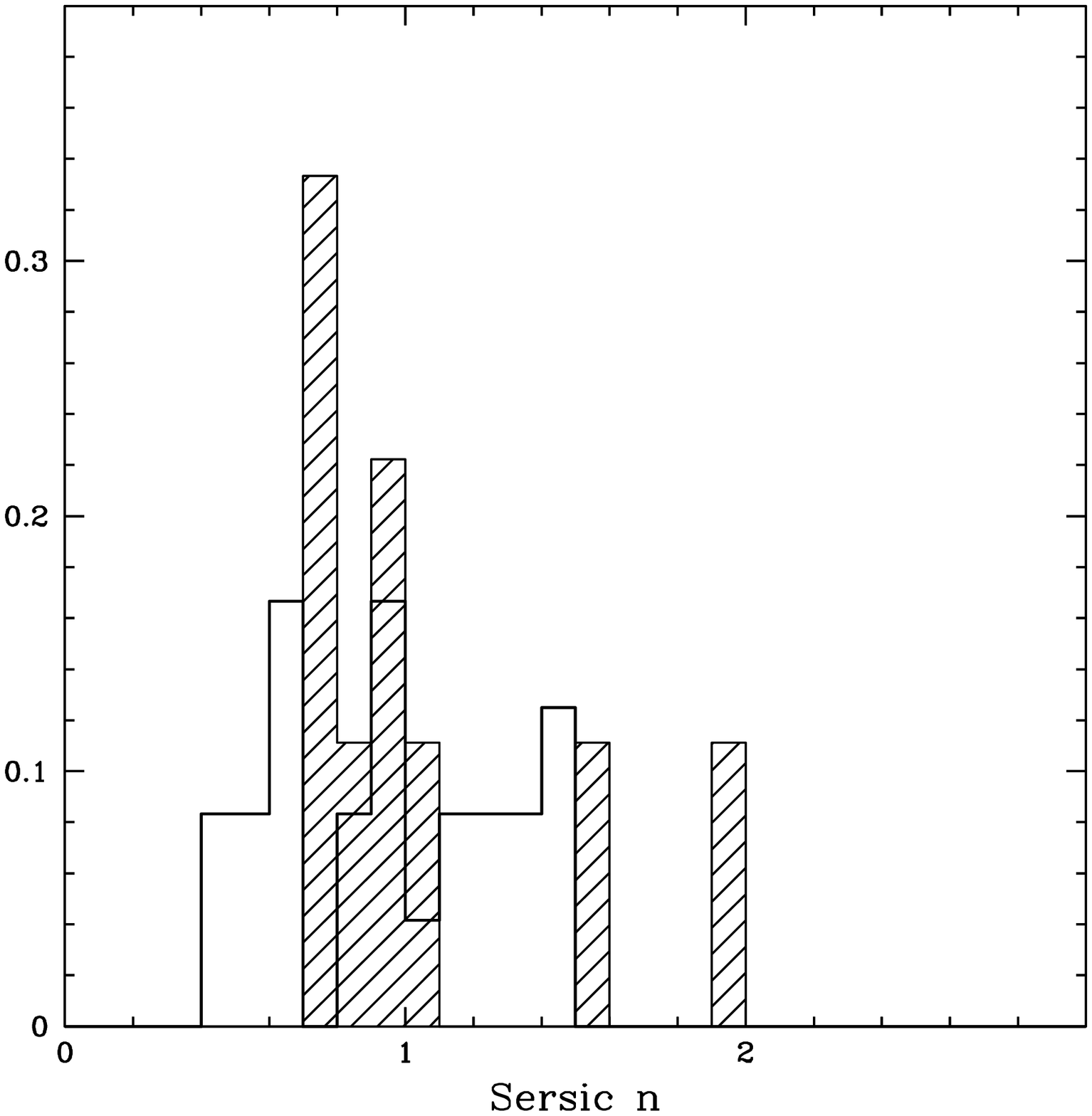}
\caption{Distributions of bulge Sersic indices and half-light radii in arcseconds for our
  galaxies (dashed histogram) and the galaxies of MacArthur et al.\ (2004).
\label{fig:struct1}
}
\end{center}
\end{figure}

\begin{figure}
\begin{center}
\includegraphics[scale=0.6]{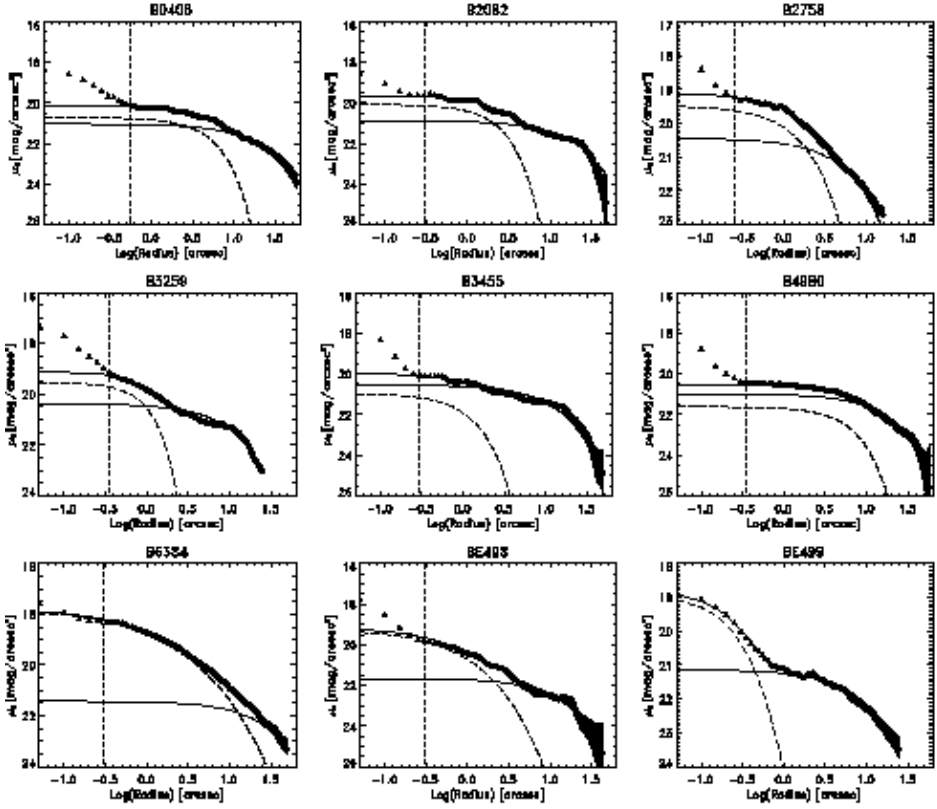}
\caption{The
  $B-$band surface brightness profiles extracted from the F435W ACS images
  ({\it triangles}) with overplotted the best-fit Sersic- bulge plus
  exponential-disk models ({\it solid lines}). The {\it dashed} lines
  represent the bulge components and the {\it dotted} lines represent the disk
  components of the best fit models. The short-dashed vertical lines indicate
  the innermost point in the surface brightness profiles which were used to
  perform the analytical fits. The excesses at small radii are contributed by the nuclear star 
  clusters discussed in Paper 2.
\label{fig:profilesB}
}
\end{center}
\end{figure}

\begin{figure}
\begin{center}
\includegraphics[scale=0.6]{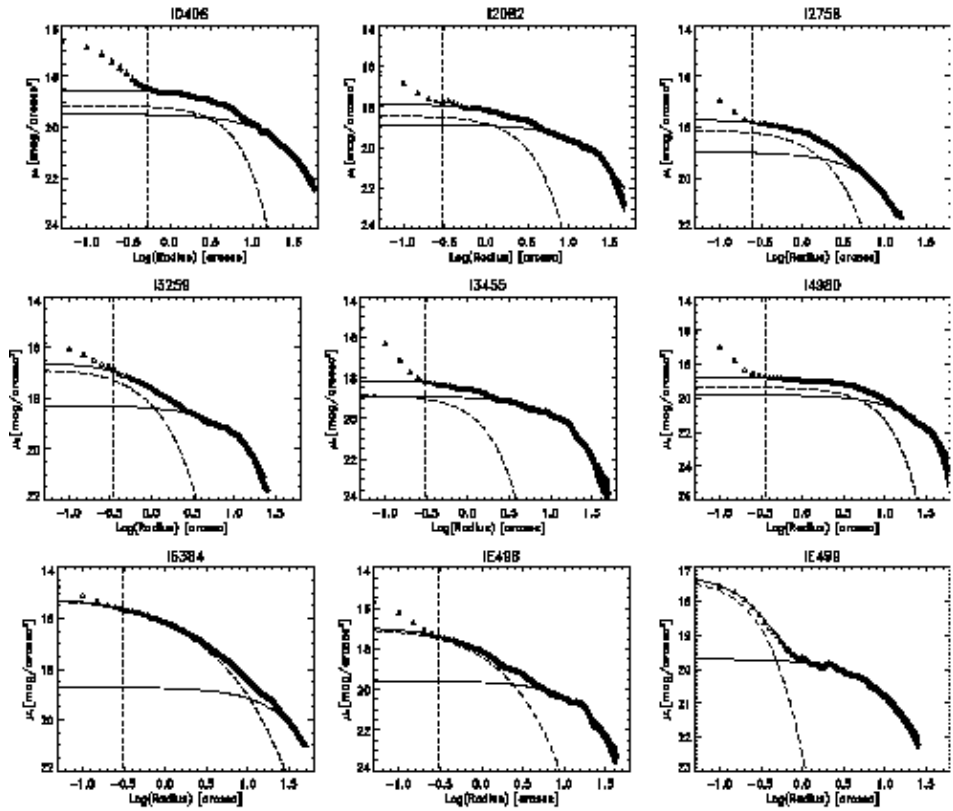}
\caption{The  $I-$band surface brightness profiles extracted from the F814W ACS images
  ({\it triangles}). Other symbols are as in Figure~\ref{fig:profilesB}.
\label{fig:profilesI}
}
\end{center}
\end{figure}

\begin{figure}
\begin{center}
\includegraphics[scale=0.6]{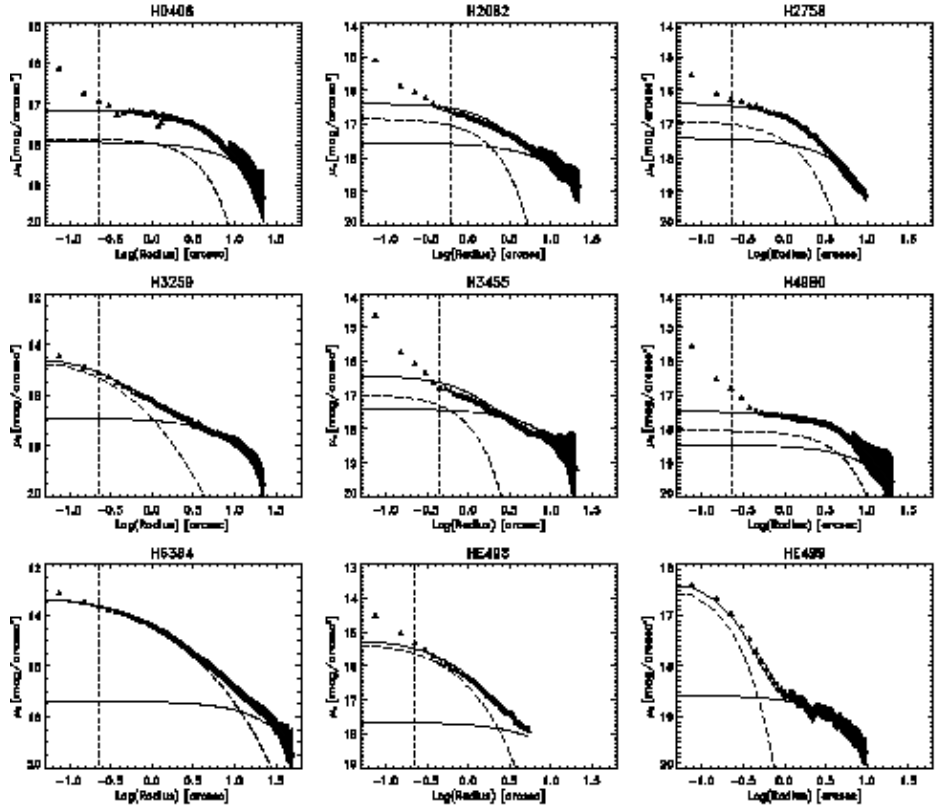}
\caption{The $H-$band surface brightness profiles
  of Carollo et al.\ (2001), obtained from the NICMOS F160W images,
  radially-extended in this work by means of ellipse-fitting to archival 2MASS
  data, when available. Other symbols are as in Figure~\ref{fig:profilesB}.
\label{fig:profilesH}
}
\end{center}
\end{figure}

A more realistic dust model has been discussed in Disney et al.\ (1989).
This model  assumes that the stars and dust have
exponential distributions in both the radial and vertical directions with
radial scale lengths $h_*$ and $h_d$, and vertical scale lengths $z_*$ and
$z_d$; the model has also been adopted by MacArthur et al.\ to discuss their
results on late-type galaxies. These authors adopt the same vertical and radial
scale length ratios for the dust and stars, and a central $V-$band optical depth
of $\tau_{V} = 1$, in agreement with multi-band surface brightness profile
modeling of massive edge-on disks and 2MASS analyses of disk galaxies
(Xilouris et al.  1999; Masters et al.\ 2003).  In Figure~\ref{fig:MIvsIH}
we plot the reddening vector for this realistic dust model as obtained using a
similar set of parameters as in McArthur et al. This reddening vector lies
rather close to parallel to the one obtained for the foreground screen model, 
indicating that our main conclusions are valid
independent of which of the two dust model distributions that we have discussed
is considered.
 
Two final considerations alleviate our concerns regarding the possible
impact of diffuse dust on our results.  First, there appears to be some
consensus that dust does not play a major role in establishing the colors in
late-type spiral galaxies. Specifically, the comparison of
optical/near-infared colors of edge-on and face-on spiral galaxies indicates
that dust plays little role in the color properties of disk galaxies with $v_c
< 120$ km s$^{-1}$ (Dalcanton \& Bernstein 2002), which is the case for the
majority of the galaxies in our sample.  Second, in order to explain as purely
dust effects the red colors of some of the bulges, or the color variations
between bulges and inner disks, an average central optical depth of $1.5$
would be required. This is significantly larger than   the value    estimated on average
for     late-type disk galaxies (Xilouris et al.\ 
1999).  Of course, future high-resolution ultraviolet and infrared data are
ultimately awaited to   confirm (or deny)   these expectations.

\newpage

\clearpage

\begin{deluxetable}{lcccccccc}
\tabletypesize{\scriptsize}
\tablecaption{Basic parameters for the sample galaxies.\label{tbl-1}}
\tablewidth{0pt}
\tablehead{
\colhead{Name} & 
\colhead{$\alpha$(J2000)} & 
\colhead{$\delta$(J2000)} & 
\colhead{$B_T$} & 
\colhead{D} & 
\colhead{Type} &
\colhead{E($B-V$)} &
\colhead{M$_H$} &
\colhead{$v_{\rm max}$} \\ 
\colhead{} & 
\colhead{($^h$ $^m$ $^s$)} & 
\colhead{($^{\deg}$ $'$ $''$)} &
\colhead{(mag)} & 
\colhead{(Mpc)} & 
\colhead{} & 
\colhead{(mag)} &
\colhead{(mag)} & 
\colhead{(km s$^{-1}$)} \\
\colhead{[1]} & 
\colhead{[2]} & 
\colhead{[3]} & 
\colhead{[4]} & 
\colhead{[5]} & 
\colhead{[6]} & 
\colhead{[7]} &
\colhead{[8]} &
\colhead{[9]} \\
}
\startdata
NGC~0406   & 01 07 24.10& $-$69 52 35.0 &13.10 &23 &SAS5*/Sc    &  0.024 &$-21.31\pm0.05$ &$104.7\pm 3.3$\\
NGC~2082   & 05 41 51.20& $-$64 18 04.0 &12.62 &19 &SBR3/SBb    &  0.057&$-21.85\pm    0.04$&$47.5\pm 3.4$\\
NGC~2758   & 09 05 30.80& $-$19 02 38.0 &13.46 &30 &PSB.4P?/Sbc &  0.129&$-20.40\pm0.10$ &$87.1\pm 2.4$\\
NGC~3259   & 10 32 34.68& $+$65 02 26.8 &12.97 &27 &SXT4*/SBbc  &  0.011& $-21.73\pm0.02$&$109.1\pm 7.8$\\
NGC~3455   & 10 54 31.20& $+$17 17 02.8 &12.87 &19 &PSXT3/...   &  0.027&$-20.77\pm0.04$& $92.2\pm 6.6$\\
NGC~4980   & 13 09 10.20& $-$28 38 28.0 &13.19 &22 &SXT1P?/SBa  &  0.077&$-20.87\pm0.05$& $69.7\pm 4.3$\\
NGC~6384   & 17 32 24.42& $+$07 03 36.8 &11.14 &26 &SXR4/SBbc   &  0.150&$-24.47\pm0.03$&$167.5\pm 5.0$\\
ESO~0498G5 & 09 24 41.10& $-$25 05 33.0 &13.96 &37 &SXS4P/SBbc  &  0.107&$-22.22\pm0.04$&         -\\          
ESO~0499G37& 10 03 42.10& $-$27 01 39.0 &13.15 &14 &SXS7*/SBc   &  0.075&$-17.82\pm0.08$&$80.3\pm 4.9$\\
\enddata
\tablecomments{Right ascension, Declination, and total blue magnitude
  are from the RC3 catalog (deVaucouleurs et al.\ 1991).
  Morphological classification are from the RC3 ({\it left}) and from
  the UGC ({\it right}) catalogs. The color excess $E($B-V$)$ is taken
  from Schlegel et al. (1988).  The galaxy absolute $H$-band magnitude M$_H$ is computed using the 2-MASS observed magnitude and the listed distance.
  The last column lists $v_{\rm max}$ from
 Paturel et al.\ (2003), except for NGC~2082 for which $v_{\rm max}$ was
  derived from $W_{20}$ (\hi\ line velocity width at 20\% of the
  maximum) taken from Kilborn et al.\ (2002).}
\end{deluxetable}

\begin{deluxetable}{lcccccc}
\tabletypesize{\scriptsize}
\tablecaption{Structural parameters from the Sersic fits of the bulge component.\label{tbl-2a}}
\tablewidth{0pt}
\tablehead{
\colhead{ } &
\colhead{r$_0$-r} &
\colhead{m$_{B,{\rm bulge}}$} &
\colhead{m$_{I,{\rm bulge}}$} &
\colhead{m$_{H,{\rm bulge}}$} &
\colhead{$r_{\rm e,I}$}&
\colhead{Sersic $n$}  \\
\colhead{Name} &
\colhead{($''$)} &
\colhead{(mag)} &
\colhead{(mag)} &
\colhead{(mag)} &
\colhead{($''$)} &
\colhead{}\\
}
\startdata
NGC~0406    & 0.55--57& 15.61$\pm$ 0.05  & 14.02$\pm$0.04  & 12.72$\pm$0.05  &  5.75$\pm$ 0.06   &  0.73$\pm$   0.07 \\
NGC~2082    & 0.30--47& 16.51$\pm$ 0.01  & 14.90$\pm$0.04  & 13.24$\pm$0.03  &  2.74$\pm$ 0.18   &  0.80$\pm$   0.08 \\
NGC~2758    & 0.25--16& 16.21$\pm$ 0.03  & 14.93$\pm$0.03  & 13.81$\pm$0.08  &  2.41$\pm$ 0.08   &  0.96$\pm$   0.07 \\
NGC~3259    & 0.35--25& 18.04$\pm$ 0.10  & 15.48$\pm$0.05  & 13.37$\pm$0.01  &  1.20$\pm$ 0.11   &  1.04$\pm$  0.10 \\
NGC~3455    & 0.30--50& 18.95$\pm$ 0.06  & 16.89$\pm$0.07  & 15.10$\pm$0.03  &  1.33$\pm$ 0.34   &  0.80$\pm$  0.24 \\
NGC~4980    & 0.35--57& 15.36$\pm$ 0.03  & 13.74$\pm$0.05  & 12.41$\pm$0.07  &  7.33$\pm$ 0.18   &  0.76$\pm$   0.06 \\
NGC~6384    & 0.30--50& 13.22$\pm$ 0.05  & 11.18$\pm$0.06  &  9.49$\pm$0.03  &  9.16$\pm$ 0.01   &  1.96$\pm$  0.01 \\
ESO~498G5   & 0.30--43& 16.74$\pm$ 0.06  & 14.66$\pm$0.01  & 13.03$\pm$0.04  &  2.20$\pm$ 0.03   &  1.55$\pm$   0.03 \\
ESO~499G37  & 0.05--25& 19.75$\pm$ 0.06  & 18.28$\pm$0.02  & 16.84$\pm$0.08  &  0.31$\pm$ 0.01   &  0.96$\pm$   0.04 \\
\enddata
\tablecomments{Best--fit parameters for the bulge 
 component of each galaxy. Specifically, listed are the
bulge half--light radius $r_{e,\rm I}$, the bulge total 
apparent magnitude $m_{I,B,H{\rm bulge}}$, and the bulge Sersic index $n$. 
The range of radii used
for the fits are listed in the second column. }
\end{deluxetable}

\begin{deluxetable}{lccccc}
\tabletypesize{\scriptsize}
\tablecaption{Structural parameters from the exponential fits of the disk components.\label{tbl-2b}}
\tablewidth{0pt}
\tablehead{
\colhead{ } &
\colhead{r$_0$-r} &
\colhead{$h_{I}$}&
\colhead{$\mu_{0\,B, {\rm disk}}$}&
\colhead{$\mu_{0\,I, {\rm disk}}$}&
\colhead{$\mu_{0\,H, {\rm disk}}$}\\
\colhead{Name} &
\colhead{($''$)} &
\colhead{($''$)} &
\colhead{(mag/"$^2$)} &
\colhead{(mag/"$^2$)} &
\colhead{(mag/"$^2$)} \\
}
\startdata
NGC~0406    & 0.55--57&  21.30$\pm$0.70 & 21.09$\pm$  0.02 &19.48$\pm$0.01 & 17.93$\pm$  0.12\\
NGC~2082    & 0.30--47&  15.97$\pm$0.36 & 20.88$\pm$  0.03 &18.86$\pm$0.03 & 17.55$\pm$  0.13\\
NGC~2758    & 0.25--16&   6.44$\pm$0.36 & 20.57$\pm$  0.08 &18.96$\pm$0.15 & 17.41$\pm$  0.12\\
NGC~3259    & 0.35--25&   9.35$\pm$0.12 & 20.37$\pm$  0.01 &18.28$\pm$0.01 & 16.90$\pm$  0.05\\
NGC~3455    & 0.30--50&  12.38$\pm$0.24 & 20.55$\pm$  0.01 &18.90$\pm$0.01 & 17.40$\pm$  0.06\\
NGC~4980    & 0.35--57&  16.98$\pm$1.80 & 21.17$\pm$  0.08 &19.80$\pm$0.30 & 18.46$\pm$  0.04\\
NGC~6384    & 0.30--50&  23.73$\pm$0.42 & 21.43$\pm$  0.40 &18.71$\pm$0.03 & 17.37$\pm$  0.85\\
ESO~498G5   & 0.30--43&  13.34$\pm$0.54 & 21.75$\pm$  0.03 &19.57$\pm$0.03 & 17.66$\pm$  0.70\\
ESO~499G37  & 0.05--25&  10.36$\pm$0.27 & 21.11$\pm$  0.01 &19.70$\pm$0.01 & 18.55$\pm$  0.07\\
\enddata
\tablecomments{Best--fit parameters for the
disk components of each galaxy.  The listed parameters are the disk scale length $h_{I}$, 
and the central surface brightness $\mu_{0;I,B,H}$. The range of radii used
for the fits are listed in the second column.}
\end{deluxetable}

\begin{deluxetable}{lcc}
\tabletypesize{\scriptsize}
\tablecaption{Color difference between bulges and disks.\label{tbl-col}}
\tablewidth{0pt}
\tablehead{
\colhead{} & 
\colhead{$\Delta(B-I)$} & 
\colhead{$\Delta(I-H)$}\\
\colhead{Name} & 
\colhead{(mag)} & 
\colhead{(mag)}\\
}
\startdata
NGC~0406    &  0.211 &   -    \\
NGC~2082    &$-$0.128&  0.325    \\
NGC~2758    &   0.35 & -   \\
NGC~3259    &  0.479 &  0.607    \\
NGC~3455    &  0.366 &  0.418    \\
NGC~4980    &  0.522 &   -    \\
NGC~6384    &  0.092 &  0.239    \\
ESO~498G5   &  0.252 &   -     \\
ESO~499G37  &  0.249 &  0.357    \\
\enddata

\tablecomments{Color differences between the 
  bulge colors, derived using the bulge-disk photometric decompositions  based, for all bands, on the shapes of the $I$-band analytical fits to the surface brightness profiles, and the
  colors of the disks, as estimated at a galactocentric distance of 5
  bulge half-light radii. The colors are corrected for   Galactic   extinction.  }
\end{deluxetable}

\begin{deluxetable}{lcccc}
\tabletypesize{\scriptsize}
\tablecaption{Stellar populations of late-type bulges: SSP models.\label{tbl-4}}
\tablewidth{0pt}
\tablehead{
\colhead{Name} & 
\colhead{$Age$} & 
\colhead{Z} & 
\colhead{$(M/L)_{H}$} &
\colhead{Bulge Mass} \\
\colhead{} & 
\colhead{(Gyr)} & 
\colhead{($Z_{\odot}$)} & 
\colhead{($M_{\odot}$/$L_{\odot}$)} &
\colhead{($M_{\odot}$)}\\ 
}
\startdata
NGC~0406      &1.5  & 0.75 & 0.65 & $6.4\times 10^8$  \\
NGC~2082$^{a}$&0.5  & 2.0  & 0.36 & $1.5\times 10^8$  \\
NGC~2758      &1.0  & 0.4  & 0.58 & $3.5\times 10^8$  \\
NGC~3259$^{b}$&0.5  & 2.0  & 0.36 & $2.7\times 10^8$  \\
NGC~3455$^{c}$&1.0  & 2.0  & 0.48 & $3.6\times 10^7$  \\
NGC~4980      &1.5  & 1.0  & 0.70 & $8.4\times 10^8$  \\
NGC~6384      & 2.0 & 2.0  &0.70& $1.7\times 10^{10}$\\
ESO~498G5     & 3.0 & 1.5  & 0.71 & $1.3\times 10^9$  \\
ESO~499G37    & 1.0 & 1.5  & 0.54 & $4.4\times 10^6$  \\
\enddata
\tablecomments{The first two columns list the best fit age and the metallicity
  of the bulge stellar population derived using the BC01 SSP models. The
 remaining columns list the  the mass-to-light ratio in the $H$ band
  $(M/L)_H$, and the mass of the bulge. \\
  $^a$ 
  Assumed $A_V=0.5$. \\
  $^b$ Assumed $A_V=1.5$. \\
  $^c$ Assumed $A_V=0.55$.
  }
\end{deluxetable}

\begin{deluxetable}{lcccc}
\tabletypesize{\scriptsize}
\tablecaption{Stellar populations of late-type bulges: ESFH models.\label{tbl-5}}
\tablewidth{0pt}
\tablehead{
\colhead{Name} & 
\colhead{$\tau$} & 
\colhead{Z} & 
\colhead{\% $M_{12-9}$} & 
\colhead{\% $M_{3-0}$} \\
\colhead{} & 
\colhead{(Gyr)} & 
\colhead{($Z_{\odot}$)} & 
\colhead{($M_{\odot}$)} &
\colhead{($M_{\odot}$)} \\ 
}
\startdata
NGC~0406    & 5.6  &   0.35  & 47 &    9 \\
NGC~2082    & 7.6  &   2.00  & 41 &   12 \\
NGC~2758    & 500  &   0.20  & 25 &   25 \\
NGC~3259$^a$& 4.3  &   2.50  & 53 &    7 \\
NGC~3455    & 3.3  &   2.25  & 61 &    4 \\
NGC~4980    & 5.6  &   0.35  & 47 &    9 \\
NGC~6384    & 3.3  &   1.50  & 61 &    4 \\
ESO~498G5   & 2.6  &   1.25  & 69 &    2 \\
ESO~499G37  & 12.  &   1.00  & 35 &   16 \\
\enddata
\tablecomments{Metallicity $Z$ and $e-$folding time $\tau$ describing
  the best fit ESFHs of the galaxies. The last two columns list the fraction of
  mass formed between 12 and 9 Gyrs ago, and in the last 3 Gyrs,
  respectivelly. \\ $^a$ Assumed $A_V=0.75$.}
\end{deluxetable}

\begin{deluxetable}{lccc}
\tabletypesize{\scriptsize}
\tablecaption{Details of the observations.\label{tbl-App1}}
\tablewidth{0pt}
\tablehead{
\colhead{ } & 
\colhead{t$_{\rm f330w}$} &
\colhead{t$_{\rm f435w}$} & 
\colhead{t$_{\rm f814w}$} \\ 
\colhead{} & 
\colhead{(s)} & 
\colhead{(s)} & 
\colhead{(s)} \\ 
}
\startdata
NGC~0406    &$2\times 2820$ &$2\times 1170$ & $2\times 1170$ \\
NGC~2082    &$2\times 2760$ &$2\times 1140$ & $2\times 1140$ \\
NGC~2758    &$2\times 2520$ &$2\times 1050$ & $2\times 1050$ \\
NGC~3259    &$2\times 2760$ &$2\times 1000$ &$2\times 1000$,$1\times 200$  \\
NGC~3455    &$2\times 2520$ &$2\times 1050$ &$2\times 1050$  \\
NGC~4980    &$2\times 2580$ &$2\times 1080$ &$2\times 1080$  \\
NGC~6384    &$2\times 2580$ &$2\times 1050$ &$2\times 1050$  \\
ESO~498G5   &$2\times 2640$ &$2\times 1080$ &$2\times 1080$  \\
ESO~499G37  &$2\times 2640$ &$2\times 1080$ & $2\times 1080$ \\
\enddata

\tablecomments{Number of exposures and exposure times of the ACS
  observations in the filters F330W, F435W, and F814W.
All the observations were taken
  with gain 2 with both HRC--F330W and WFC--F814W images, and with gain 1 for
  WFC--F435W (except for NGC~3259 which was observed with gain$=2$).
  }
\end{deluxetable}

\end{document}